\documentclass[oneside, a4paper, onecolumn, 11pt, authoryear]{elsarticle} 

\usepackage[left=2cm,top=2cm,bottom=1.5cm,right=2cm]{geometry}
\usepackage{hyperref}
\usepackage[utf8]{inputenc}
\usepackage{graphicx} 		
\usepackage{amsmath}  		
\usepackage{natbib}	
\usepackage{eurosym}
\usepackage{caption} 
\usepackage{gensymb} 
\usepackage{booktabs}
\usepackage{array,multirow}
\usepackage{siunitx}
\usepackage{setspace}

\usepackage{lineno}

\usepackage{lingmacros}
\usepackage{tree-dvips}
\usepackage[utf8]{inputenc}
\usepackage[english]{babel}
\usepackage{amsfonts}
\usepackage{subcaption}
\usepackage{listings}
\usepackage{float}
\usepackage{verbatim}
\usepackage{color}

\usepackage{enumitem}
\setitemize{noitemsep,topsep=0pt,parsep=0pt,partopsep=0pt,leftmargin=*}
\usepackage{amssymb}

\usepackage{nopageno}
\usepackage{enumitem}

\usepackage{fancyhdr}
\newenvironment{itemize*}%
  {\begin{itemize}%
    \setlength{\itemsep}{0pt}%
    \setlength{\parskip}{0pt}}%
  {\end{itemize}}

\usepackage{enumitem}

\usepackage{tabularx}

\begin{document}

\begin{center}

\LARGE{\textbf{Wave measurements from ship mounted sensors in the Arctic marginal ice zone \\ ~}}

~ \\

\large{Trygve. K. Løken$^{a,}$\footnotemark , Jean Rabault$^{b,a}$, Atle Jensen$^{a}$, Graig Sutherland$^{c}$, Kai H. Christensen$^{b}$ and Malte Müller$^{b}$}
 
\footnotetext{E-mail address and phone number for corresponding: trygvekl@math.uio.no (+47) 94885767 (T.K. Løken)\\ 
E-mail addresses: jean.rblt@gmail.com (J. Rabault), atlej@math.uio.no (A. Jensen),\\ graigory.sutherland@canada.ca (G. Sutherland), kaihc@met.no (K.H. Christensen), maltem@met.no (M. Müller).} 

\end{center}\

\noindent $^{a}$ Department of Mathematics, University of Oslo, Oslo, Norway \\
$^{b}$ Norwegian Meteorological Institute, Oslo, Norway \\
$^{c}$ Environment and Climate Change Canada, Québec, Canada \\

\section*{Abstract}

This study presents wave measurements in the marginal ice zone (MIZ) obtained from ship mounted sensors. The system combines altimeter readings from the ship bow with ship motion correction data to provide estimated single point ocean surface elevation. Significant wave height and mean wave period, as well as one-dimensional wave spectra are derived from the combined measurements. The results are compared with integrated parameters from two spectral wave models over a period of eight days in the open ocean, and with spectra and integrated parameters derived from motion detecting instruments placed on ice floes inside the MIZ. Mean absolute errors of the integrated parameters are in the range $13.4$-$29.9$\% when comparing with the spectral wave models and $1.0$-$9.6$\% when comparing with valid motion detecting instruments. The spatial wave damping coefficient is estimated by looking at the change in spectral wave amplitude found at discrete frequency values as the ship was moving along the longitudinal direction of the MIZ within time intervals where the wave field is found to be approximately constant in time. As expected from theory, high frequency waves are effectively dampened by the presence of sea ice. The observed wave attenuation rates compare favourably with a two-layer dissipation model. Our methodology can be regarded as a simple and reliable way to collect more waves-in-ice data as it can be easily added to any ship participating to ice expeditions, at little extra cost. \\

\textbf{Keywords:} Sea ice dynamics, wave measurements, marginal ice zone, wave attenuation

\section{Introduction} \label{intro}

Sea ice is a major feature in the polar environments, it has importance for Arctic ecosystems as well as for global ocean and atmospheric circulation. A decline in the Arctic ice cover has been observed over the past decades \citep{feltham2015arctic}. Interactions between sea ice and surface gravity waves play an important role in breakup and reduction of ice cover. The decline in ice cover leads to a larger fetch where waves build up more energy and enhance the breakup and melting process in a positive feedback mechanism \citep{thomson2014swell}. The mixture of icebergs, floes and grease ice found in the interface between solid ice, such as land fast ice or pack ice, and the open ocean, is called the marginal ice zone (MIZ). Previous studies have found that high frequency wind waves are effectively dampened in the MIZ \citep{weber1987wave,wadhams1988attenuation}, which reduces the ice cover break up rate.     

Recent changing conditions in the polar regions have allowed for increased human activities, which raises the importance of better forecast models and improved physical understanding of the environment to ensure safe operations \citep{fritzner2019impact}. In situ wave measurements can increase our understanding of global climate systems and provide data for calibration of numerical models. Also, mathematical models to describe wave attenuation in ice, e.g. \cite{weber1987wave}, \cite{newyear1997comparison} and \cite{sutherland2019two}, need experimental data for validation and improvement. However, experimental data are relatively sparse due to the inaccessibility of the regions where sea ice is present, combined with the harsh and dangerous environment for both researchers and instruments \citep{squire2007ocean}. \ 

We present here results from shipborne wave measurements in the MIZ. This system has not up until now been used deep into the MIZ, therefore, the novelty of our observations. The methodology, first described in \cite{christensen2013surface}, combines a bow mounted altimeter and a motion correction device. We provide wave spectra and integrated parameters from spectra, which are important quantities when considering wave-ice interactions. Significant wave height and mean periods are compared with a spectral wave forecast model over a period of eight days in the open ocean. We also compare measurements in the MIZ with data from wave measuring instruments consisting of inertial motion units (IMUs) placed on ice floes \citep{rabault2019open}. From the spectra, the spatial damping coefficient is found as a function of wave frequency, which can be compared to attenuation models.  \ 

In this paper, the data acquisition and processing methods are described in Section~\ref{method}. The results are presented in Section~\ref{results}, which is divided in two parts. A comparison with spectral models and in situ instruments is outlined in Section~\ref{section:dataValidation}. In Section~\ref{damping_section}, we present results on wave attenuation and the spatial damping coefficient is compared to the theoretical model of \citep{sutherland2019two}. Finally, a discussion follows in Section~\ref{discussion} and the concluding remarks are given in Section~\ref{conclusions}.

\section{Data and Methods} \label{method}

The data were obtained during a research campaign in the Barents Sea with R/V Kronprins Haakon in September 2018 as part of the Nansen Legacy project \citep{reigstad2017nansen}. The vessel is $100$~m long and with a beam of $21$~m. In total, the cruise lasted two weeks, during which continuous measurements were made. The results from the MIZ presented here were recorded on September 19 when the ship ventured approximately $28$~km into the MIZ. Four stops were made to deploy in situ waves-in-ice (WII) instruments on ice floes \citep{rabault2019open}, which were used to validate our ship mounted system. Upon the return to the open ocean, a total of seven stops were made with more or less equal spacing on a close to straight south-southwest heading where measurements for wave damping estimates were carried out. Twenty minutes samples were recorded as the ship was freely drifting on each station. The location, starting time and wave travel distance (WTD) for each measurement in the MIZ are summarized in Table~\ref{ship_stops}. Prefix $1$ denotes the four stops into the MIZ while prefix $2$ denotes the seven stops out of the MIZ. \

\begin{table}[h]
\centering 
\begin{tabular}{c c c c}  
\toprule
Stop & Time & Position (N/E) & WTD [km] \\ [0.5ex]
\midrule
1.1 & 04:22 & 82.126/20.736 &  \\[0.5ex]
1.2 & 06:28 & 82.246/20.245 &  \\[0.5ex]
1.3 & 08:59 & 82.355/19.803 &  \\[0.5ex]
1.4 & 12:20 & 82.436/19.674 &  \\[0.5ex]
2.1 & 13:11 & 82.421/19.579 & 92.7 \\[0.5ex]
2.2 & 14:32 & 82.359/19.544 & 69.2 \\[0.5ex] 
2.3 & 15:40 & 82.294/19.389 & 58.3 \\[0.5ex]
2.4 & 16:54 & 82.228/19.275 & 20.8 \\[0.5ex]
2.5 & 18:00 & 82.163/19.183 & 6.7 \\[0.5ex]
2.6 & 19:08 & 82.099/19.046 & 0 \\[0.5ex]
2.7 & 20:09 & 81.994/18.982 & 0 \\[0.5ex]
\bottomrule
\end{tabular}
\caption{Time and location where measurements were carried out inside the MIZ. Prefix $1$ for stop number indicate deployment of WII instruments while prefix $2$ indicate measurement of damping coefficient. WTD through the MIZ are listed for the stops with prefix $2$.} 
\label{ship_stops} 
\end{table}

\begin{figure}[H]
\begin{subfigure}[H]{0.5\linewidth}
\includegraphics[width=8cm]{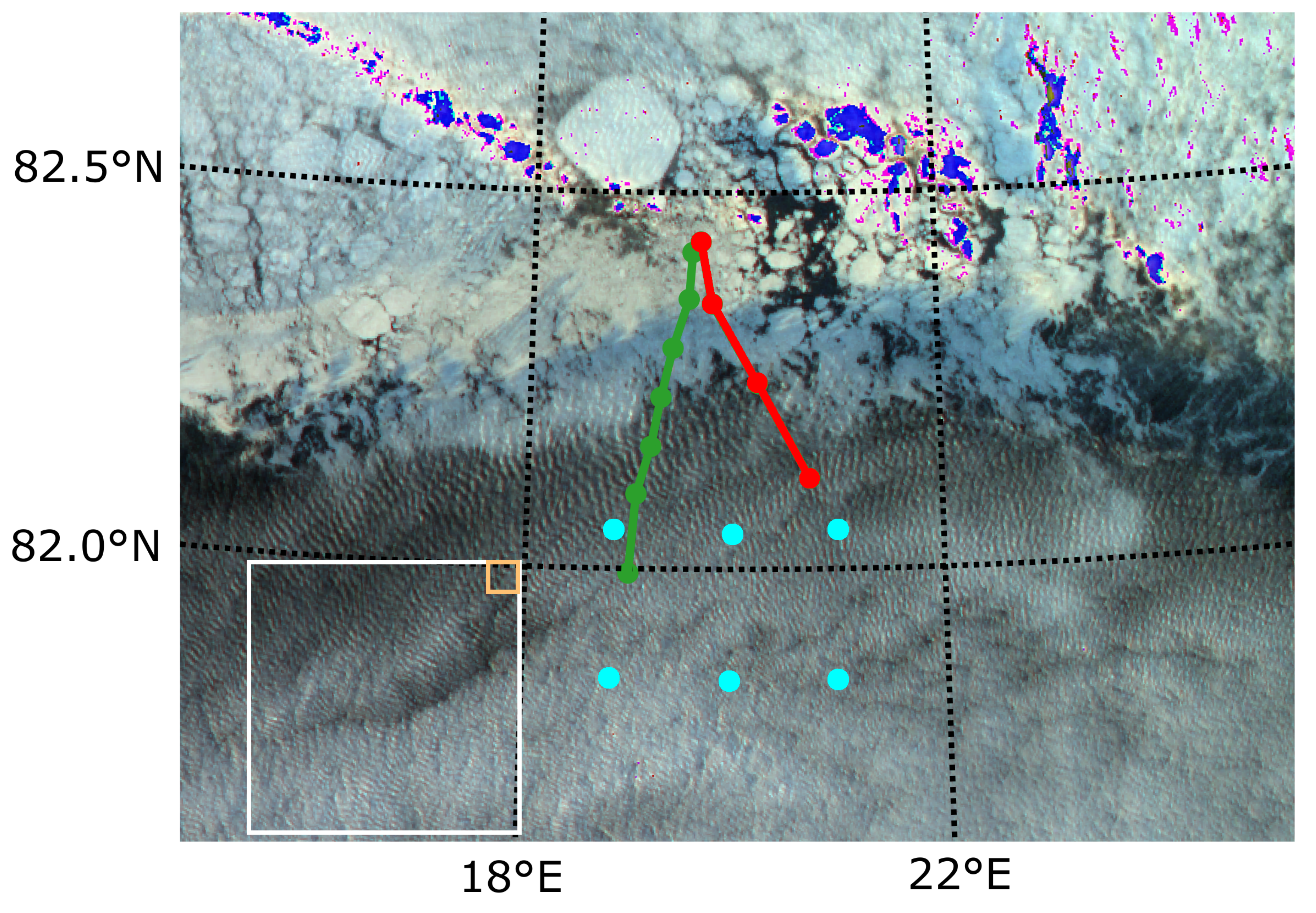} 
\caption{Satellite image 14:55 UTC 19-09-2018.}
\label{fig:satellite}
\end{subfigure}
\hfill
\begin{subfigure}[H]{0.5\linewidth}
\includegraphics[width=8cm]{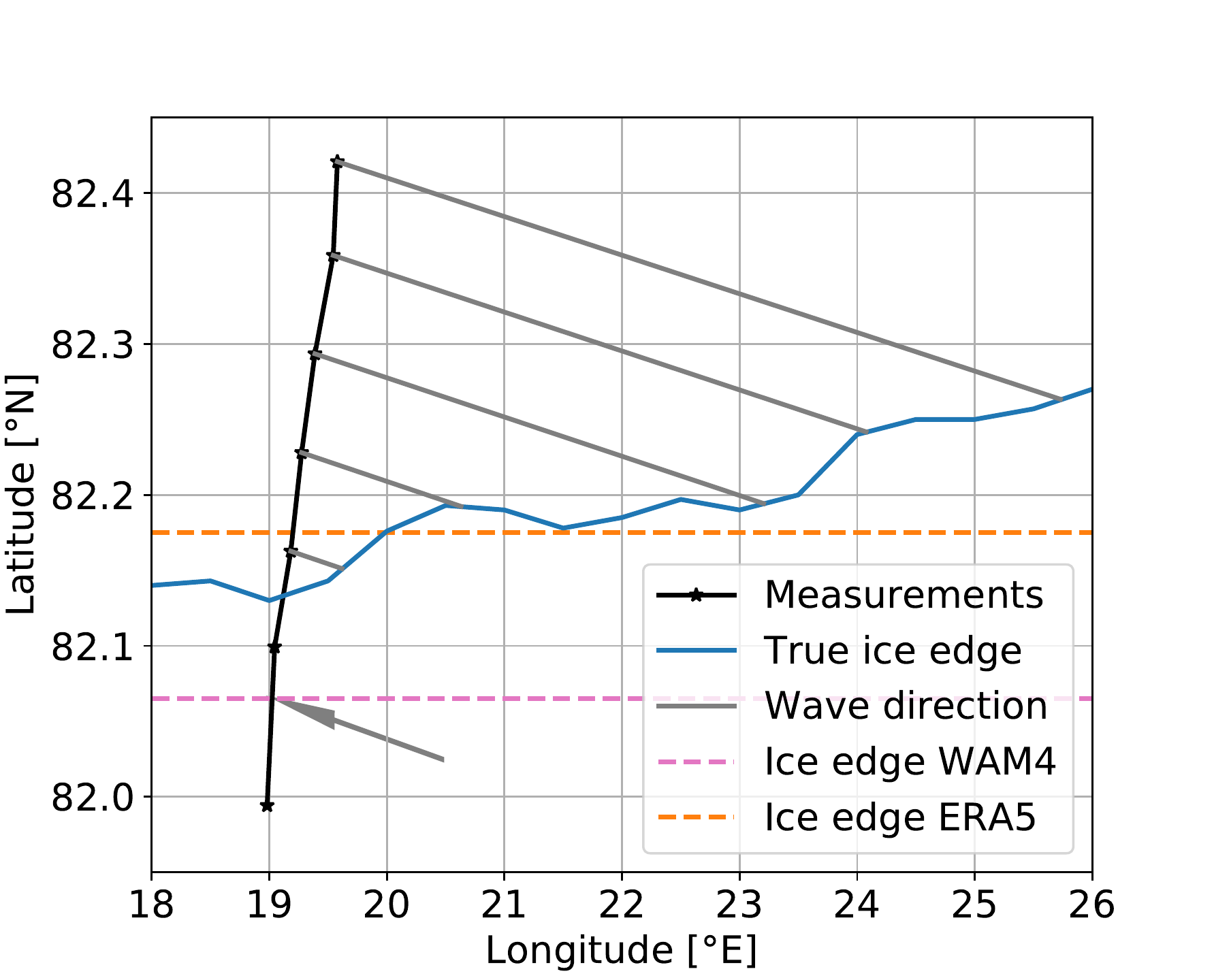} 
\caption{Wave attenuation measurements.}
\label{fig:trajectory}
\end{subfigure}
\hfill
\caption{MIZ and location of measurements. a) Satellite image with ship trajectory into (red) and out of (green) the MIZ. Cyan dots mark the location where wave parameters from the spectral model WAM-4 are extracted to check for constant sea state. White and orange squares indicate the horizontal resolution of the spectral wave models ERA5 and WAM-4 respectively. Source: MODIS Corrected Reflectance (250~m resolution, sharpened) b) Wave direction from WAM-4 and WTD through the ice for each measurement (used for wave damping coefficient). The ice edge found from the satellite image (blue) and the one used in WAM-4 (pink) and ERA5 (orange) are indicated.}
\label{fig:test}
\end{figure}

Figure~\ref{fig:satellite} is a MODIS Corrected Reflectance satellite image, which shows the ship trajectory into (red) and out of (green) the MIZ and the ice edge at approximately $82.2$~\degree N below the thin cloud cover. The ice concentration for the four stops $1.1$-$1.4$ was estimated to be roughly $10$\%, $30$\%, $90$\% and $100$\% respectively in \cite{rabault2019open}. From Fig.~\ref{fig:satellite} the ice edge is roughly located by visual inspection along the relevant longitudes and recreated in Fig.~\ref{fig:trajectory} (blue) along with the location of the wave attenuation measurements, i.e. stops $2.1$-$2.7$ (black). This figure also indicates the ice edge used in the spectral models WAM-4 (pink) and ERA5 (orange), which are further described in Section~\ref{wave_model}, and the mean wave direction on September 19 (gray) estimated by WAM-4. We use the "going-to" convention for wave direction and ship heading and the "coming-from" convention for wind direction and define directions as clockwise rotation from the geographic north. The gray lines illustrate the WTD through the ice for each stop. \

\subsection{Data acquisition} \label{data_acq}

\begin{figure}[H]
\begin{subfigure}[H]{0.5\linewidth}
\includegraphics[width=8cm]{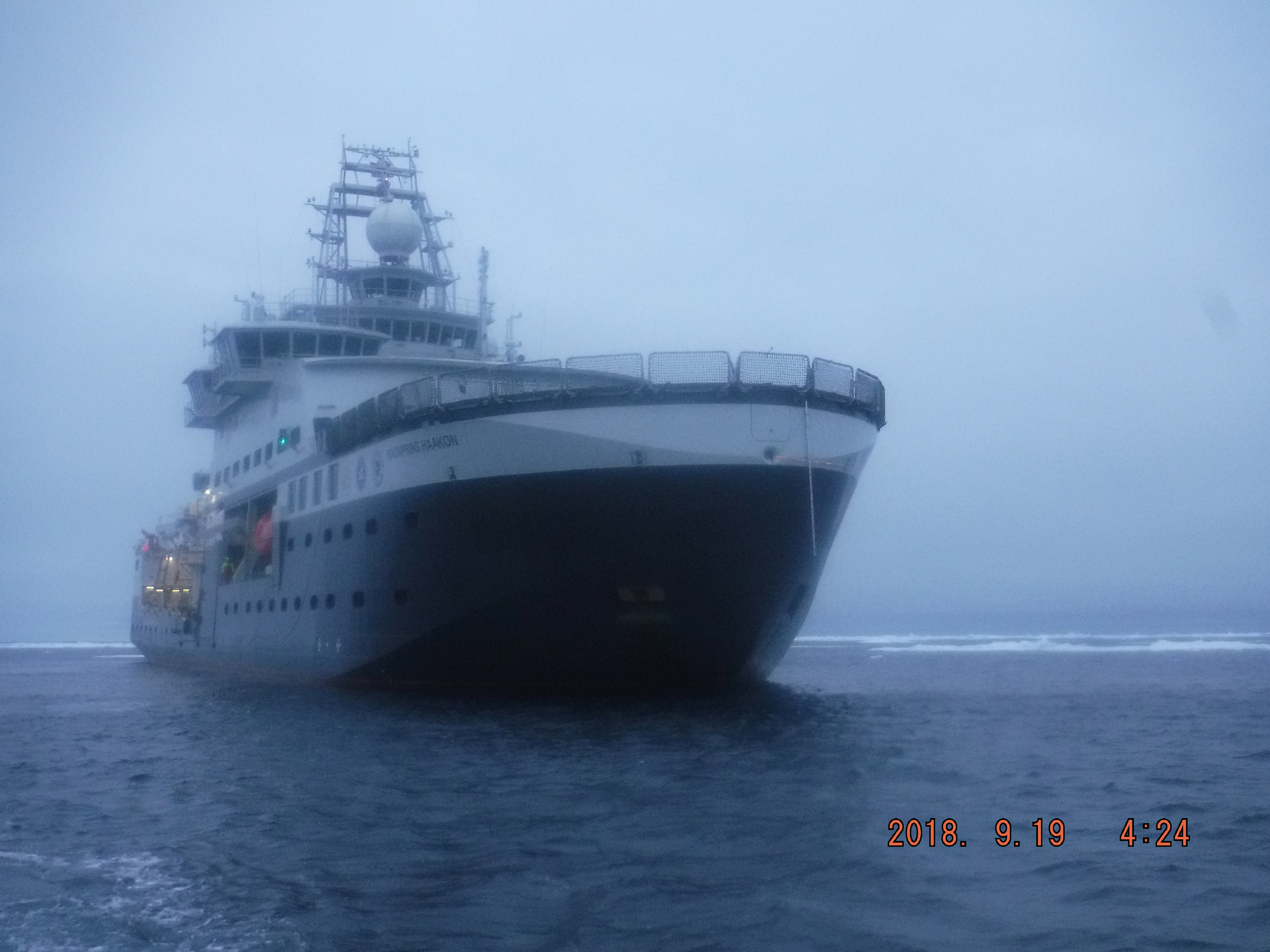} 
\caption{UG seen from the front.}
\label{fig:ship}
\end{subfigure}
\hfill
\begin{subfigure}[H]{0.5\linewidth}
\includegraphics[width=8cm]{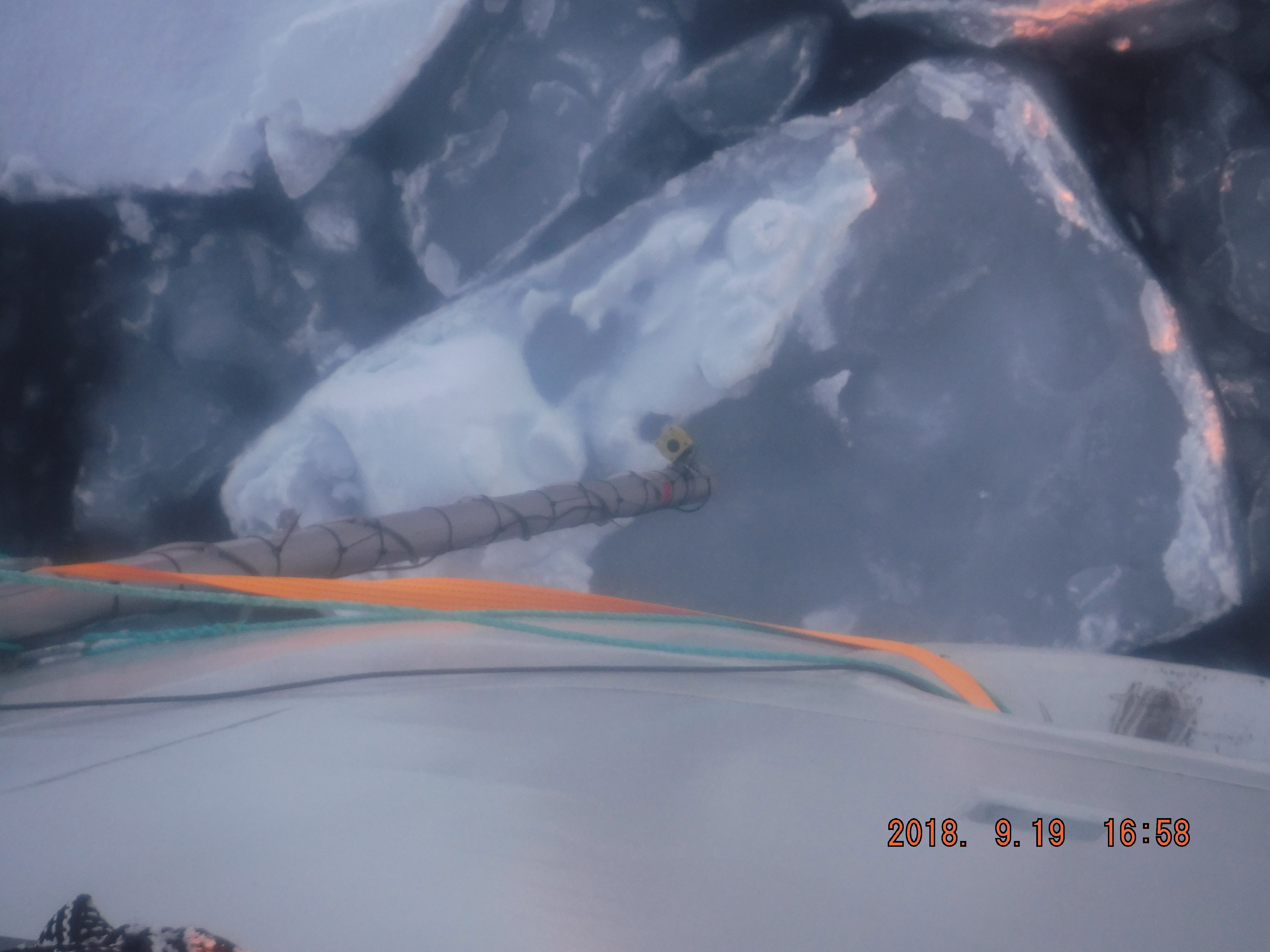} 
\caption{UG seen from above.}
\label{fig:pole}
\end{subfigure}
\hfill
\caption{Installation of the downward facing UG on a rigid pole in the ship bow.}
\label{fig:test}
\end{figure}

The instrument setup consisted of an ultrasonic gauge (UG) that measured ocean surface elevation relative to the ship bow. Figure~\ref{fig:ship}~and~\ref{fig:pole} shows the downward facing UG mounted on a rigid pole. Estimated absolute surface elevation was obtained after correcting for ship motion by means of an IMU placed on deck, also in the bow section of the ship. The horizontal and vertical distances between the UG and the IMU were $x_{lev}=2.5$~m and $y_{lev}=6$~m respectively. \

We used a UG (Banner QT50ULB) with approximately $0.2$-$8$~m range. The instrument emits $75$~kHz ultrasonic pulses at a $10.4$~Hz sampling rate. Distance $L$ between the UG and the ocean surface is calculated internally from the time delay of the echo, where the speed of sound in air is temperature compensated with an integrated thermometer. Mean $L$ was approximately $5$~m and the effective beam width, i.e. the diameter of the circular area where the signal was reflected on the ocean surface, was approximately $0.9$~m. This footprint is very low compared to the wavelength $\lambda = 156$~m of a typical $0.1$~Hz ocean wave, as the one shown in Fig.~\ref{time_series}. The wavelength is found from $\lambda=2\pi/k$, and the wavenumber $k$ from the linear deep water dispersion relation:

\begin{equation}
\label{dispersion}
\omega^{2} = gk,
\end{equation}

\noindent where $\omega = 2\pi f$ is the angular frequency, $f$ the wave frequency and $g$ the acceleration due to gravity. The depth dependency is neglected here since the water depth was $\approx 3500$~m in the region. The use of (\ref{dispersion}) in the MIZ can be justified by the fact that the dispersion relation in ice deviates little from the deep water dispersion relation in the frequency band where wave motion is present \citep{marchenko2017field}. \

A feature of the UG called Auto-Window was enabled, meaning that a $1$~m sensing window was centered around a taught length. The taught length $\bar{L}$ was found automatically by time averaging $L$ over a couple of wave periods before each measurement was initiated. We later display the output signal of the instrument as $D=L-\bar{L}$ in Figures~\ref{time_series}, \ref{fig:floe} and \ref{fig:saturated_time}. The Auto-Window feature was chosen out of practical reasons and to increase instrument accuracy at the expense of a reduced range ($|D|<0.5$~m), because the intention was initially to only measure in the Barents Sea MIZ where low wave heights are normally expected. However, it was decided to measure continuously during the whole cruise, also in the open ocean where the waves were generally larger than in the MIZ. Consequently, the instrument range was exceeded (saturated) at times with large wave amplitudes. Note that saturation only occured when $|D|>0.5$~m, not necessarely when the trough-to-crest amplitude exceeded $1$~m, depending on ship responce. For future expeditions we would recommend to pre-calibrate the UG to allow for a larger sensing window if rougher waves are expected, for example in the Suthern Ocean MIZ, to reduce the probability of saturation. With the right instrument calibration, the measurement range could rather be limited by the actual instrument range and the mean vertical distance between the instrument and the surface ($0.2$-$8$ and $5$~m respectively in our case). \cite{christensen2013surface} reports an equivalent setup of the system to be useful for significant wave heights up to $4$~m. \ 

\begin{figure}
  \begin{center}
    \includegraphics[width=.55\textwidth]{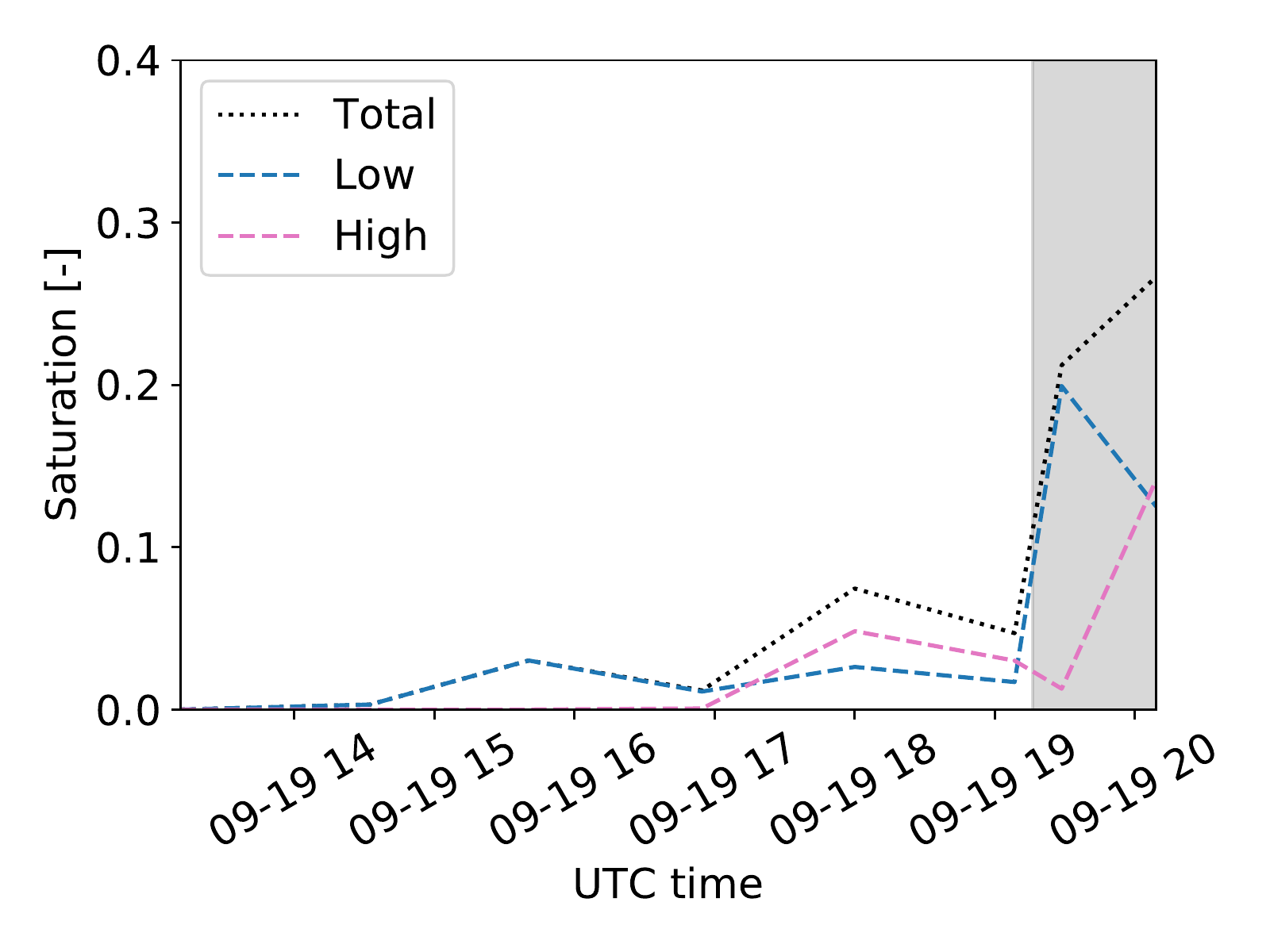} 
    \caption{\label{saturation_prop} Saturation proportion (i.e. proportion of time when the UG range was exceeded) for stops $2.1$-$2.7$, where "total" saturation proportion is the sum of "low" and "high" saturation proportion. Area of discarded data due to total saturation exceeding $10$\% threshold is shaded.}
  \end{center}
\end{figure}

We define a low saturation when $D<-0.5$~m and a high saturation when $D>+0.5$~m. In these cases, the output signal of $D$ is simply $\pm 0.5$~m, although the distance is actually smaller or greater, and this can be seen as a "cut" graph in Fig.~\ref{fig:saturated_time}. We define the saturation proportion as the ratio of time where saturation occurs over total sample time. Figure~\ref{saturation_prop} shows the saturation proportion of each sample for stops $2.1$-$2.7$, where "total" saturation proportion is the sum of "low" and "high" saturation proportion. We accept up to $10$\% total saturation and will therefore discard the sample recorded at stop $2.7$, which had a total saturation of approximately $20$\%, in our wave attenuation analysis in Section~\ref{damping_section}. \

The system can measure trough-to-crest amplitudes up to $1$~m before the range of the UG is exceeded. This constraint sets an upper limit for how large significant wave height the instrument theoretically can record in a time series without exceeding the measurement range. Significant wave height ($SWH$) from time series is defined as:

\begin{equation}
\label{SWH}
SWH = 4\sigma,
\end{equation}

\noindent where $\sigma$ is the standard deviation of the surface elevation $\eta$. In the simplified case of a pure sine wave, the trough-to-crest amplitude is equal to $2\sqrt{2}\sigma$. In this idealized example where ship motion is not considered, the limiting upper value of $SWH$ is approximately $1.4$~m for the UG. \

As motion correction device, we used an IMU (VectorNav VN100) similar to the ones in \cite{rabault2019open}. It features 3-axis accelerometers and 3-axis gyroscopes measuring at a rate of $800$~Hz. After an internal Kalman filtering, the instrument gives an output frequency of $80$~Hz. The gyros yield rotation angles about all three axis directly. Vertical acceleration is integrated twice to obtain ship vertical displacement about the mean, where positive values are displacement above the mean. Details on the integral scheme, data filtering and other technical information on the instrument can be found in \cite{rabault2019open}. \

In order to obtain time series of the surface elevation, UG and IMU data at the same time instance were needed. We solved this by defining a common sampling rate of $10$~Hz, which should be sufficient for resolving all relevant ocean surface features. All data were then interpolated on the common time base for the analysis. \

\subsection{Ship response}

In general, if the wavelength $\lambda$ is large when compared with the diameter $d$ of a floating body, the body will tend to follow the path of a fluid particle at the free surface. In the opposite case when $\lambda/d$ is small, the waves are losing their influence on the behavior of the body. The response amplitude operator (RAO) of a vessel (or any object) is the ratio of response amplitude over incoming wave amplitude. RAOs in all modes for R/V Kronprins Haakon are reported in \cite{ytterland2016motion}. RAOs depend on the wave heading angle $\beta$, which is defined as the relative angle between the ship heading and wave direction. Hence, $\beta = 0 \degree$ corresponds to following seas, $\beta = 90 \degree$ corresponds to beam seas, and $\beta = 180 \degree$ corresponds to head seas. \

\subsection{Motion correction}

Downwards facing altimeters mounted on fixed structures are common for measuring ocean surface elevation $\eta$, which is a function of time \citep{reistad2011high, karin2013andrea}. Correction is required when the altimeter is mounted on a floating structure like a ship. The parameter $\eta$ is a function of the position in the horizontal plane in addition to time when the ship is free to move, but we assume the horizontal drift velocity to be much smaller than the phase speed of the waves, so that the dependency of position can be neglected. We define a coordinate system with the $(x,y,z)$ axis to be aligned horizontally in the direction from stern to bow, vertically in upward direction and horizontally in the direction from port to starboard, respectively. The origin coincides with the IMU. Along axis translation are surge, heave and sway while rotation about the axis are roll, yaw and pitch, respectively. Only heave $\xi$, roll angle $\phi_{R}$ and pitch angle $\phi_{P}$ affect the distance $L$ measured by the UG and need to be addressed when compensating. \

In the (imaginary) case where the ship lies completely still, as if it was fixed in position, surface elevation will be given by $\eta = \bar{L} - L$. When heave response is present, we simply subtract the vertical displacement $\xi$, which is positive when the ship is above its mean level and negative when the ship is below. When rotation about the horizontal axis is included, two aspects need to be considered. First, the UG measures the distance to the surface with an angle about the vertical axis. The true vertical distance is obtained by multiplying $L$ with the cosine of both $\phi_{R}$ and $\phi_{P}$ as shown in the second term on the R.H.S. of (\ref{UG}). Second, there is the lever effect due to the horizontal and vertical distance between the $UG$ and the $IMU$, $x_{lev}$ and $y_{lev}$ respectively. The UG is displaced in vertical direction relative to the ocean surface when the UG rotates around the IMU about the $x$-axis and the $z$-axis, which corresponds to ship roll and pitch respectively. We name these displacements roll and pitch elevation effects, and define them respectively as $y_{lev}(1-cos(\phi_{R}))$, which is always positive, and $x_{lev}sin(\phi_{P})$, which is positive when $\phi_{P}$ is positive and negative otherwise. \

All corrections are combined and we obtain an expression for the ocean surface elevation:     

\begin{equation}
\label{UG}
\eta = \bar{L} - Lcos(\phi_{R})cos(\phi_{P}) - \xi - y_{lev}(1-cos(\phi_{R})) - x_{lev}sin(\phi_{P}),
\end{equation}  

\noindent which is similar to Eq. 1 in \cite{christensen2013surface} except from the roll and pitch elevation effects in the fourth and fifth term on the R.H.S. of (\ref{UG}), which are new considerations. \ 

\begin{figure}
  \begin{center}
    \includegraphics[width=.55\textwidth]{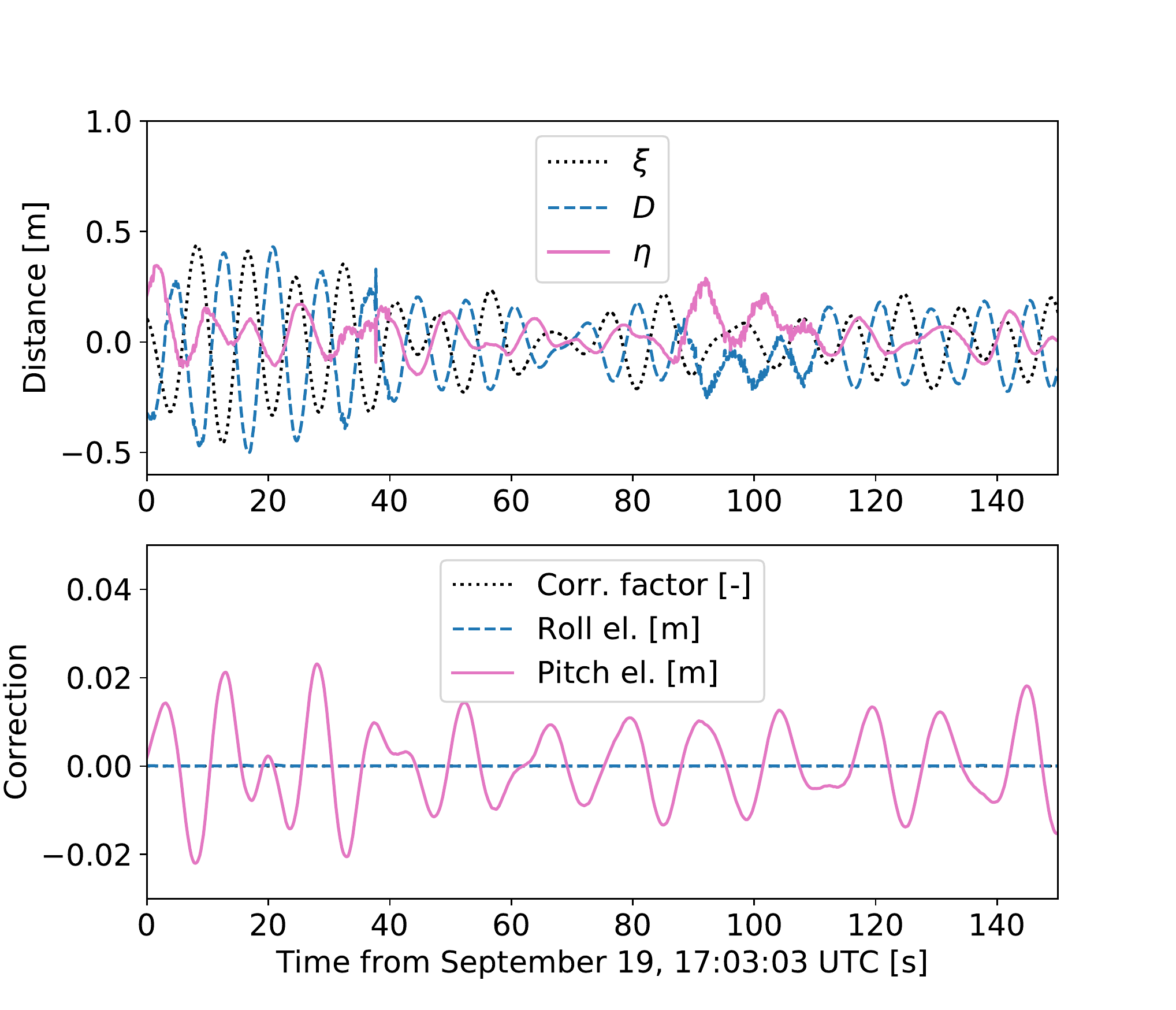} 
    \caption{\label{time_series} $150$~s time series of UG and IMU measurements from stop $2.4$. In upper panel: Ship heave response $\xi$, UG-reading $D$ and surface elevation $\eta$. In lower panel: UG angle correction factor (corresponds to the second term on R.H.S. of (\ref{UG}), displayed with black dotted line), roll and pitch elevation effect (fourth and fifth term on R.H.S. of (\ref{UG}) respectively).}
  \end{center}
\end{figure}

A typical example of a $150$~s time series inside the MIZ is shown in Fig.~\ref{time_series}. Trough-to-crest amplitudes up to approximately $0.4$~m and frequencies of around $0.1$~Hz are found in this specific sample, as seen in the upper panel. Ship heave response $\xi$ and UG reading $D=L-\bar{L}$ are also presented in the upper panel and are in the same order of magnitude as the surface elevation. The lower panel shows roll and pitch elevation effects (fourth and fifth term on R.H.S. of (\ref{UG}) respectively) in addition to the UG angle correction factor corresponding to the second term on R.H.S. of (\ref{UG}), here presented as $1 - cos(\phi_{R})cos(\phi_{P})$. These results indicate that the roll elevation effect and the UG angle correction factor are $\mathcal{O}(10^{-3})$ relative to the wave amplitude and thus negligible in this case. The pitch elevation effect is on the other hand approximately $0.02$~m and $\mathcal{O}(10^{-1})$ relative to the wave amplitude and is therefore important to include in the processing.    

\begin{figure}[H]
\begin{subfigure}[H]{0.5\linewidth}
\includegraphics[width=8cm]{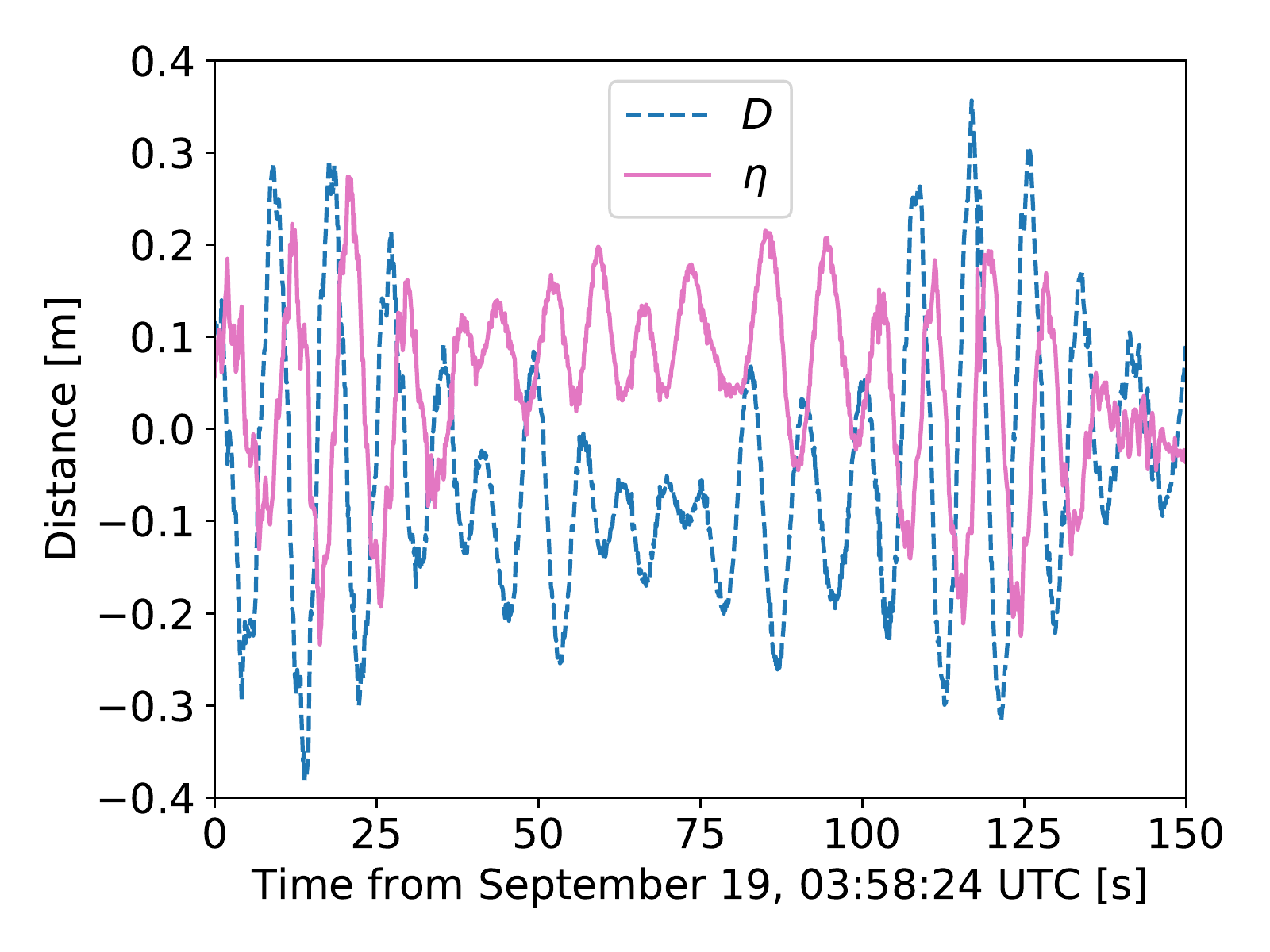} 
\caption{Surface elevation over ice floe.}
\label{fig:floe}
\end{subfigure}
\hfill
\begin{subfigure}[H]{0.5\linewidth}
\includegraphics[width=8cm]{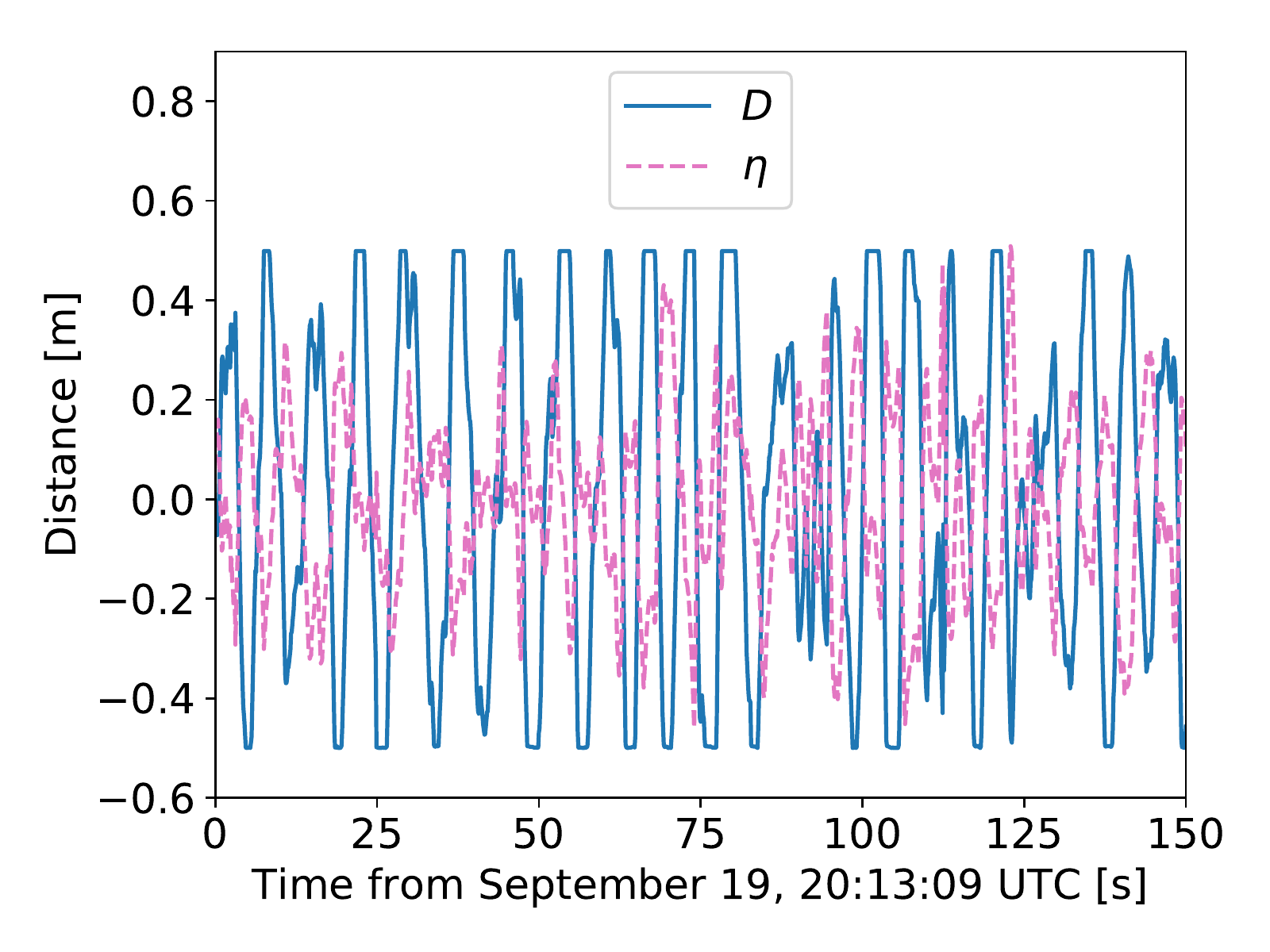} 
\caption{Saturated UG reading.}
\label{fig:saturated_time}
\end{subfigure}
\hfill
\caption{Time series of UG-reading $D$ and surface elevation $\eta$. a) Stop $1.1$ where an ice floe most likely drifts under the sensing window of the instrument. b) Stop $2.7$ with approximately $20\%$ saturated UG signal.}
\label{fig:test}
\end{figure}

A section of the time series from stop number $1.1$ is presented in Fig.~\ref{fig:floe}, where the mean value of the surface elevation $\eta$ jumps from zero up to about $0.1$~m after roughly $35$~s and then jumps back to zero at roughly $100$~s. This step is most likely caused by an ice floe drifting under the sensing window of the UG for a short period, which for example also occurs at stop number $2.4$ as illustrated in Fig.~\ref{fig:pole}, and demonstrates the UG's ability to receive reflected signals off an ice cover. The crest-to-trough amplitude is approximately $0.4$~m for the waves and $0.1$~m for the step presented in Fig.~\ref{fig:floe}. If the step is interpreted as a long wave, its frequency is less than $0.008$~Hz, which is much lower than the lower cut-off frequency $f_{min}$ explained in Section~\ref{section:stats}, and should therefore not affect the spectral analysis presented in this study. \ 

An example of a saturated sample can be seen in Fig.~\ref{fig:saturated_time}. This time series is extracted from stop number $2.7$. The UG reading $D$, is clearly flattened out at the peaks where the range of the instrument is exceeded. Total saturation for this sample was approximately $20$\%. \ 

\subsection{Statistical parameters and spectrum} \label{section:stats}

We obtain surface elevation from (\ref{UG}). Power spectrum density $PSD(f)$ of the surface elevation is obtained with the Welch method \citep{earle1996nondirectional} where the samples are subdivided in $q$ consecutive segments and ensemble averaged. Segment size is set to $200 s$ with $50$~\% overlap. With $20$~min sampling time at $10$~Hz common sampling rate, this gives a segment size of $2000$ sampling points and a total number of $12000$ sampling points for each measurement. A Hanning window is applied to each segment to reduce spectral leakage. \

Since we are later comparing our measurements with data from a spectral model, we find the statistical parameters from spectra. The mean wave period and the significant wave height are obtained from the spectral moments:

\begin{equation}
\label{moment}
m_{j} = \int_{f_{min}}^{f_{max}} f^{j}PSD(f)df,
\end{equation}

\noindent where the cutoff frequencies $f_{min}$ and $f_{max}$ are set to $0.04$~Hz and $1.0$~Hz respectively, which should include the most energetic ocean waves. The same cutoff frequencies are applied in the spectral model we have compared our results with. In our comparison of integrated parameters from bow instruments and WII instruments at stops $1.1$-$1.4$, we use $0.05$-$0.25$~Hz, which corresponds to the cutoff frequencies applied by the on-board processing unit of the WII instruments. The spectral moments can be used to estimate the mean ($T_{m01}$) and zero up-crossing ($T_{m02}$) periods from

\begin{equation}
\label{Tm01}
T_{m01} = \frac{m_{0}}{m_{1}},
\end{equation}

\noindent and

\begin{equation}
\label{Tm02}
T_{m02} = \sqrt{\frac{m_{0}}{m_{2}}},
\end{equation}

\noindent respectively. \

We use significant wave height $H_{s}$ from spectra for most of our analysis here. $SWH$ from time series of surface elevation (Eq. \ref{SWH}) is used for comparison in a redundancy check for the measurements, as the two methods ideally should yield the same result. $H_{S}$ is defined as:

\begin{equation}
\label{Hs}
H_{s} = 4\sqrt{m_{0}}.
\end{equation} 

Wave attenuation is dependent on frequency. It is therefore convenient to have a measure for wave amplitude as function of frequency when finding the spatial damping coefficient. We define a spectral amplitude $a$ at discrete forcing frequency $f_{0}$ as: 

\begin{equation}
\label{a}
a(f_{0}) = \sqrt{\int_{f_{0} - \Delta f}^{f_{0} + \Delta f}PSD(f)df},
\end{equation}

\noindent where $\Delta f$ is the bandwidth frequency set to $0.005$~Hz. Six discrete forcing frequencies ($0.076$-$0.128$~Hz with $0.0104$~Hz increments) are chosen based on the spectra presented in Section~\ref{damping_section}. Equation \ref{a} is similar to the definition of \cite{meylan2014situ}, used for attenuation analysis in the Antarctic MIZ. The spectral amplitude $a(f_{0})$ is not the physical wave amplitude in the classical sense, but an interpretation of the energy content in a finite frequency proportion of the PSD with unit meter. \

Confidence intervals for both spectra and the integrated parameters, significant wave height and amplitude, are calculated from the Chi-squared distribution, following the methodology presented in \cite{young1995determination}. For spectra, the total degree of freedom (TDF) is calculated as $TDF=2q$, and for significant wave height, TDF is found with Eq. \ref{TDF} below \citep{specialist2017confidence}. 

\begin{equation}
\label{TDF}
TDF = \frac{2q\left[\int_{f_{min}}^{f_{max}} PSD(f)df\right]^{2}}{\int_{f_{min}}^{f_{max}} [PSD(f)]^{2}df}.  
\end{equation}

Equation \ref{TDF} is also used when finding the confidence intervals for $a(f_{0})$, with the small modification of multiplying the bandwidth frequency $\Delta f$ to the denominator. \

We use the mean absolute percentage error (MAPE) to compare our system with either the spectral model or the WII instruments. Systematic bias is described with the mean percentage error (MPE). The error statistics are defined as:

\begin{equation}
\label{MAPE}
MAPE = \frac{100\%}{n}\sum_{t=1}^{n}\left|\frac{X_{t}-Y_{t}}{X_{t}}\right|, 
\end{equation}

\begin{equation}
\label{MPE}
MPE = \frac{100\%}{n}\sum_{t=1}^{n}\frac{X_{t}-Y_{t}}{X_{t}}, 
\end{equation}

\noindent where $Y_{t}$ are parameters obtained from bow measurements and $X_{t}$ are reference parameters.

\subsection{Wave models} \label{wave_model}

WAM-4 is a third generation spectral wave model \citep{komen1996dynamics,saetra2004potential}, run operationally by the Norwegian Meteorological Institute. Its performance has been validated against in-situ and EnviSat Radar Altimeter observations \citep{carrasco2014wam,gusdal2011wam}. Spatial and temporal resolution of the model is $4$~km and $1$ hour, respectively. It runs twice a day with a forecast period of 66 hours per run. In this study, we have concatenated forecasts from consecutive runs to provide a continuous time series. A hard ice boundary based on satellite images (ice concentration over 3/10th) is defined in the model. \

ERA5 reanalysis of global atmosphere and ocean waves is produced by the European Centre for Medium-Range Weather Forecasts (ECMWF). ERA5 combines historical observations into past forecasts using data assimilation systems to model meteorologic and oceanographic parameters. The operational wave model at ECMWF assimilates satellite synthetic aperture radar observations and altimeter wave heights into the numerical scheme of the WAM spectral wave model, which is coupled with the IFS atmospheric model to account for wind forcing \citep{janssen2003part}. The sea ice concentration (SIC) is based on coupled atmosphere ocean simulations of the IFS model combined with satellite observations \citep{p16555}, and wave parameters are not given for $SIC>50\%$. Reanalysis wave data from ERA5 have hourly output and a horizontal resolution of $40$~km.   \

No wave simulations are performed where the assumed ice cover is present, although the models still provide wind information everywhere within the domain. We have used model data from WAM-4 and ERA5 as a comparison to our wave observations outside the MIZ. The model ice edge was defined at $82.065$~\degree N and $82.175$~\degree N by WAM-4 and ERA5 respectively at the relevant longitude on September~19. The horizontal resolution of the models are indicated with a white and orange square in Fig.~\ref{fig:satellite} for ERA5 and WAM-4 respectively. From the models, we extract total mean wave direction $THQ$, $H_{S}$, $T_{m01}$, $T_{m02}$ (from WAM-4 only) and wind parameters. $THQ$ is defined as the mean of the two-dimensional wave spectrum over all frequencies and directions. The wave parameters are integrated parameters from spectra, but the raw spectra were not available. \

\subsection{Waves-in-ice instruments}

Four in situ WII instruments \citep{rabault2019open}, placed on ice floes close to the ship at the stops $1.1$-$1.4$, have been used to cross-validate our system inside the MIZ. The instruments measure waves with an integrated IMU, and they have been tested and used in a series of previous works \citep{rabault2016meas,rabault2017measurements,sutherland2016observations,marchenko2017field}. The WII instruments are autonomous and designed to be deployed for long durations. Compressed power spectra from $20$~min time series of surface elevation were sent via Iridium satellite and used as comparison with the spectra obtained from the bow measurements. We also compare the integrated parameters $H_{S}$ and $T_{m02}$. \

The WII instruments measured with five hours intervals to conserve battery power. Data from the first measurement were corrupted due to disturbances during carrying and placement on the ice and were discarded. The results presented here are from the second measurement of each instrument, i.e. five hours after the respective bow measurement. The temporal change of sea state over these five hour periods are addressed in Section~\ref{section:dataValidation}. The WII instruments drifted $2.4$~km, $1.7$~km, $3.5$~km and $2.2$~km during the five hours between first sample taken at stop $1.1$-$1.4$ respectively, and second sample. These are relatively small displacements compared to the distance between the WII instruments, which were an order of magnitude larger. Hence, the measurements with the two different systems were made at approximately the same place with five-hour time delay. \

\subsection{Attenuation modeling}

Previous field measurements indicate that waves decay exponentially in ice \citep{squire1980direct, wadhams1988attenuation}. For each frequency bin corresponding to the six forcing frequencies $f_{0}$, the spectral amplitudes $a$ are fitted to decreasing exponentials on the shape $a = Ce^{-\alpha x}$, where $C$ and $\alpha$ are estimated parameters and $x$ is wave traveling distance through the MIZ, by means of non-linear least squares. From the fitted curves, the spatial damping coefficients $\alpha$, which describe wave attenuation are determined from:

\begin{equation}
\label{alpha_o}
\frac{\partial a}{\partial x} = -\alpha a,
\end{equation}

\noindent for each frequency bin. Non-linear least squares are applied to fit a power function $\alpha = Af^{p}$ to the measured values of $\alpha$ as function of frequency $f$, where $A$ and $p$ are the best-fit parameters. The standard error in the exponent is obtained from the square root of the variance of $p$. \

Spatial damping coefficients found in the field measurements are compared to the model of \cite{sutherland2019two}, which presents a parameterization for wave dissipation that allows for a two-layer structure within the ice. The lower layer with thickness $\epsilon h_{i}$, where $h_{i}$ is the total ice thickness, is defined as a highly viscous layer where wave motion exists. The upper layer is defined as impermeable. With a no-slip boundary condition imposed on the bottom of the ice, Equation (16) of \cite{sutherland2019two} can be written as:

\begin{equation}
\label{alpha_two}
\alpha_{mod} = \frac{1}{2}\epsilon h_{i} k^2,
\end{equation}

\noindent where $k$ is the wavenumber, here calculated with (\ref{dispersion}). Total ice thickness is set to $h_{i}=1.1$~m based on visual observations from the field campaign. We use $\epsilon=0.02$ due to the large ice thickness, meaning that wave motion is assumed to exist in a small fraction of the total ice thickness. Note that (\ref{alpha_two}) is proportional to frequency to the fourth power. \

\section{Results} \label{results}

\subsection{Data comparison} \label{section:dataValidation} 

Independent in-situ wave measurements were not available for validation of the observations in the open ocean on this cruise. Therefore, we compare our results with WAM-4 and ERA5 spectral model data. It is worth mentioning that \cite{christensen2013surface} performed and documented validation of an equivalent setup of the system in open water against a moored Waverider buoy, model data from the ERA Interim Reanalysis and satellite observations from AVISO. \

\begin{figure}
  \begin{center}
    \includegraphics[width=.75\textwidth]{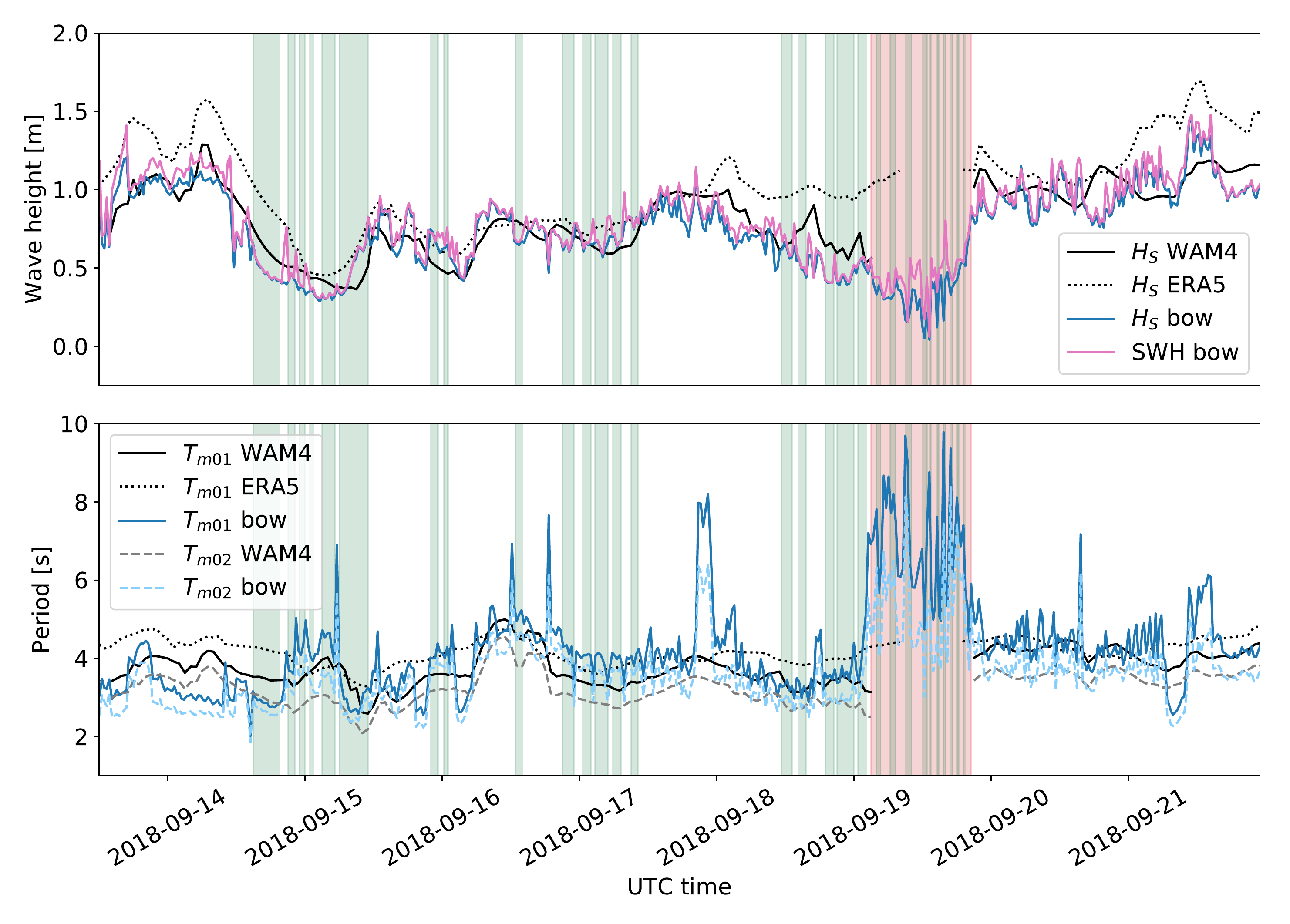} 
    \caption{\label{validation_WAM} Long time comparison between observations and the models WAM-4 and ERA5. Significant wave height from time series and spectra (upper), and periods from spectra (lower). Time periods where total saturation is below $10$\% and ship speed over ground (SOG) is below $0.5$~m/s are highlighted with green background color and the period with an ice cover (hence no model data) is highlighted with red background color. Note that ERA5 provides wave data further into the MIZ than WAM-4.}
  \end{center}
\end{figure}

Bow measurements outside the MIZ are compared with spectral model data at the ship location in Fig.~\ref{validation_WAM}. The period with no available model data due to the ice cover is highlighted with red background color and is not included in the comparison between observations and the spectral models. Valid measurements, i.e. time periods where total UG saturation is below $10$\% and ship speed over ground (SOG) is below $0.5$~m/s, are highlighted with green background color. The whole comparison spans over eight days, and periods of valid measurements are found $23$\% of the time. Error statistics are summarized in Table~\ref{table:WAM} where MAPE describes the mean absolute error and MPE describes the mean error. In general, there is a good agreement between observations and models, also for the conditions not defined as valid. The MPEs of WAM-4 do not exceed $\pm10$\% for $H_{S}$ and $T_{m01}$ in the valid periods, which indicate low systematic bias. ERA5 systematically overestimates $H_{S}$ compared to the observations, but matches the observed $T_{m01}$ better than WAM-4. A surprising result is that significant wave height has a higher MAPE in the valid periods than outside, although the difference is not large ($18.9$\% vs $13.3$\% for WAM-4 and $29.2$\% vs $21.3$\% for ERA5). MAPEs for $T_{m01}$ are about the same inside the valid periods as outside. Note that significant wave height does not exceed the theoretical limit (for sine waves) of $1.4$~m as described in Section~\ref{data_acq} within the valid data periods. MAPE between measured $H_{S}$ and $SWH$ was $11.7$\% during valid measurements (not shown in Table~\ref{table:WAM}). \

\begin{table}[h]
\centering 
\begin{tabular}{c c c c c c c}  
\toprule
\multicolumn{2}{c}{\multirow{2}{*}{Error}} & \multicolumn{3}{c}{WAM-4} & \multicolumn{2}{c}{ERA5} \\[0.5ex]
\cmidrule(lr){3-5} 
\cmidrule(lr){6-7}
\multicolumn{2}{c}{} & $H_{S}$ & $T_{m01}$ & $T_{m02}$ & $H_{S}$ & $T_{m01}$ \\ [0.5ex]
\midrule
\parbox[t]{1mm}{\multirow{2}{*}{\rotatebox[origin=c]{90}{Valid}}} & MAPE [\%] & 18.9 & 15.0 & 17.2 & 29.2 & 13.4 \\[0.5ex]
& MPE [\%] & 9.2 & -8.9 & -14.4 & 28.5 & 3.4 \\[1.5ex]
\parbox[t]{1mm}{\multirow{2}{*}{\rotatebox[origin=c]{90}{Not val.}}} & MAPE [\%] & 13.3 & 15.7 & 14.7 & 21.3 & 16.4 \\[0.5ex]
& MPE [\%] & 3.2 & -5.5 & -4.4 & 19.5 & 3.1 \\[1.5ex]
\bottomrule
\end{tabular}
\caption{Long time error statistics for significant wave height $H_{S}$ and periods $T_{m01}$ and $T_{m02}$ when comparing bow measurements with the WAM-4 and the ERA5 model. Valid periods are defined as when total UG saturation is less than $10$\% and ship speed over ground is less than $0.5$~m/s ($23$\% of the time). All other times are considered not valid. } 
\label{table:WAM}
\end{table}

\begin{figure}
  \begin{center}
    \includegraphics[width=.55\textwidth]{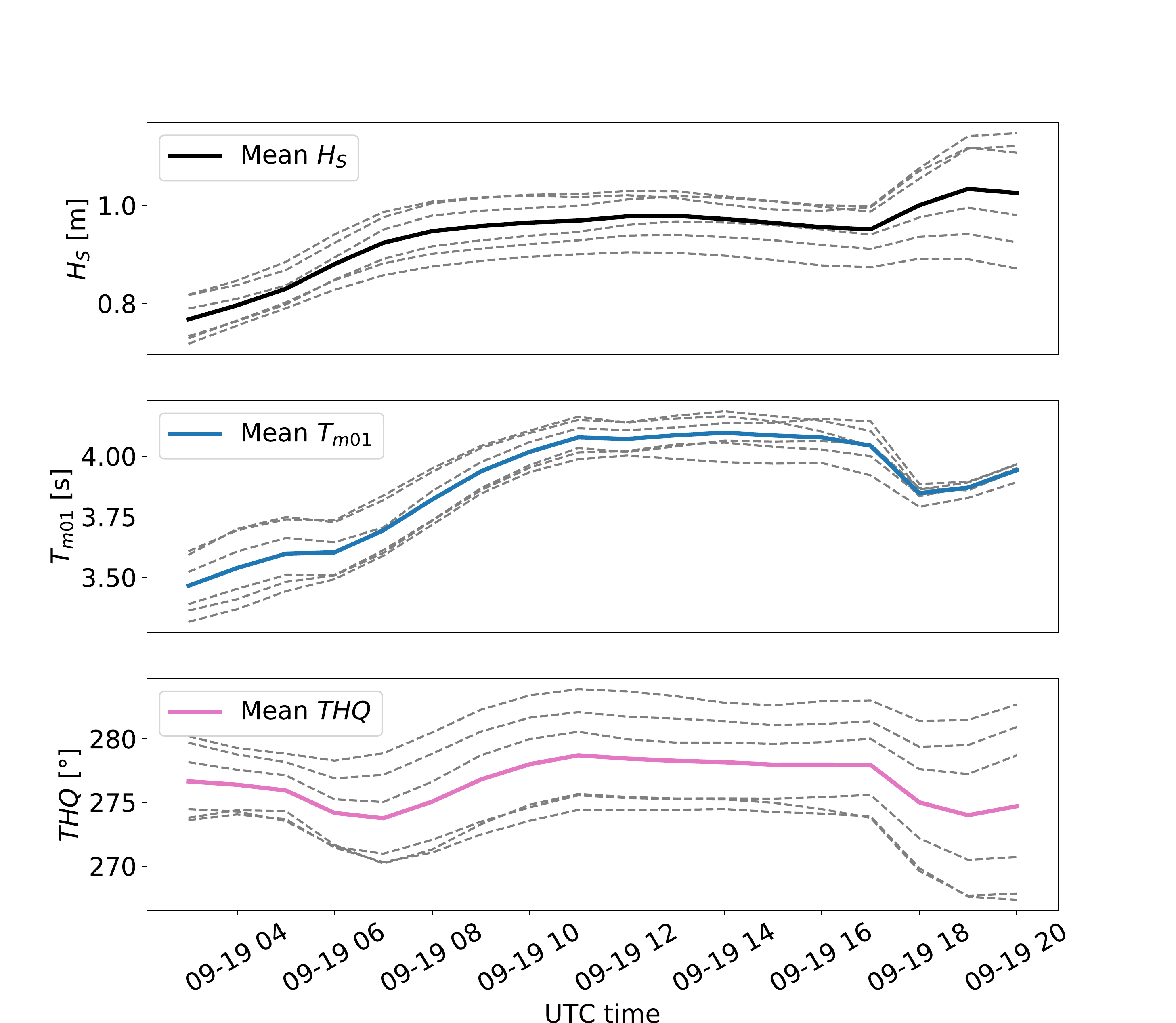} 
    \caption{\label{fig:model_outside} Time series of $H_{S}$ (upper), $T_{m01}$ (middle) and $THQ$ (lower) from WAM-4 spectral model. Values are extracted from locations indicated by the six cyan colored dots in Fig. \ref{fig:satellite} and plotted in gray. Mean values are plotted in different colors. }
  \end{center}
\end{figure}

A stationary sea state over the measurement period inside the MIZ is an advantage in our analysis. The sea state is investigated in Fig.~\ref{fig:model_outside} where time series of $H_{S}$, $T_{m01}$ and $THQ$ from WAM-4 are presented. The model data are extracted from the six points arranged in a grid configuration outside the MIZ, shown in cyan in Fig.~\ref{fig:satellite}. Time series from each grid point are plotted in gray and the mean value in a different color. All parameters were close to constant between 11:00 and 17:00 but varying with time outside of this period. This indicates a variability in sea state in the comparison of bow and WII instruments, due to the five-hour time delay between the measurements. However, the change was not dramatic. From 04:22 to 09:22 at stop $1.1$, the standard deviation in the mean samples (containing five values) were $0.06$~m, $0.10$~s and $1.00 \degree$ for $H_{S}$, $T_{m01}$ and $THQ$ respectively. From 12:20 to 17:20 at stop $1.4$, the standard deviation in the mean samples are an order lower for all parameters. The results in Fig.~\ref{fig:model_outside} also indicate a variability in sea state over the period from 13:11 to 19:08 where wave damping was measured. In this period, the standard deviations in the mean samples (containing seven values) were $0.03$~m, $0.10$~s and $1.64 \degree$ for $H_{S}$, $T_{m01}$ and $THQ$ respectively.          

\begin{figure}
  \begin{center}
    \includegraphics[width=.75\textwidth]{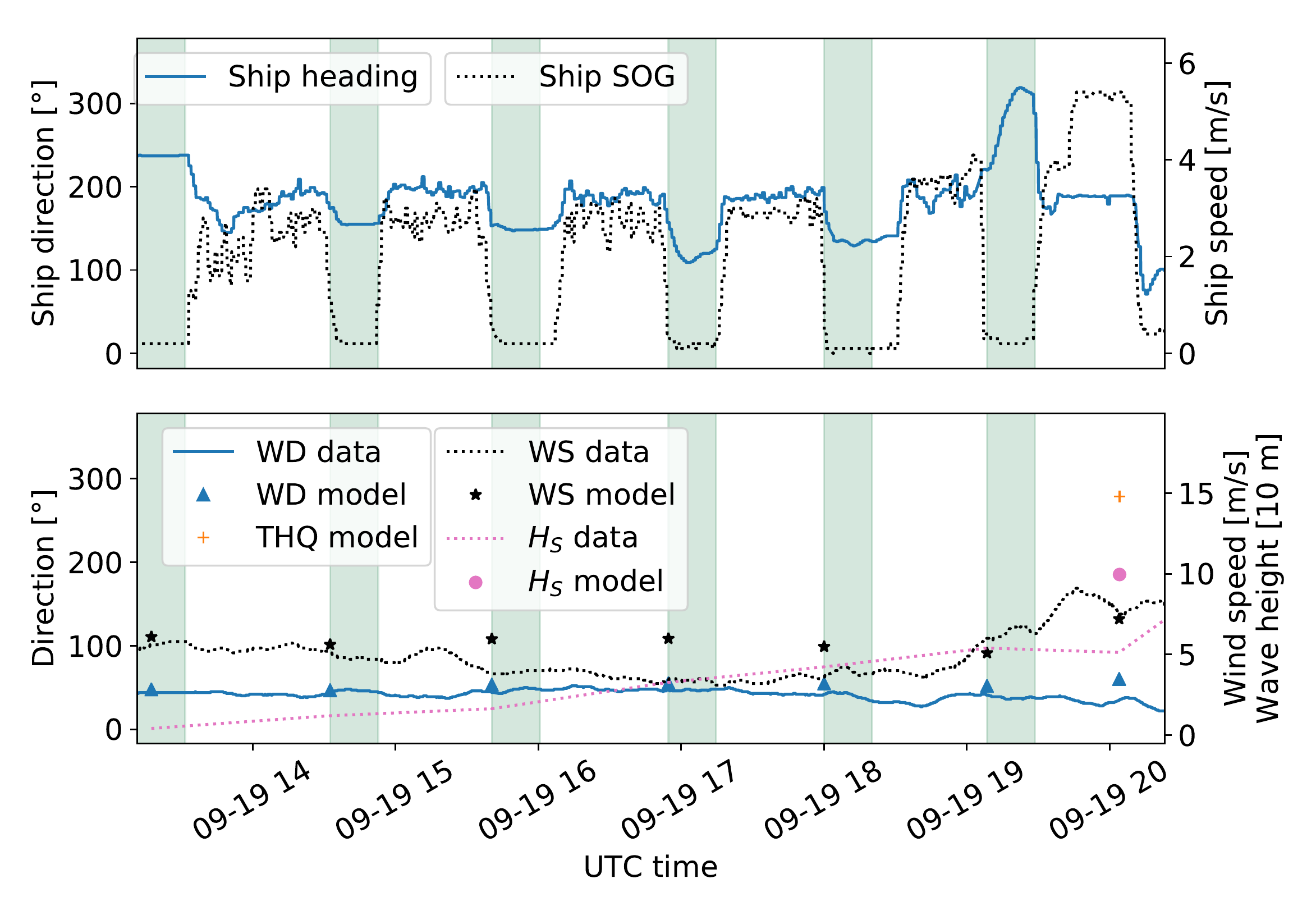} 
    \caption{\label{AWS} Comparison of observations and WAM-4 model. Ship heading and SOG (upper), and wind direction (WD) and wind speed (WS) from ship measurements and model, total mean wave direction $THQ$ from model and significant wave height $H_{S}$ (multiplied with a factor of 10) from bow measurements and model (lower). Directions increase clockwise from geographic north (zero degrees). Times with valid data are shaded.}
  \end{center}
\end{figure}

A comparison of ship measurements and available WAM-4 model data from stops $2.1$-$2.6$ is presented in Fig.~\ref{AWS}. The upper panel shows ship heading and speed over ground (SOG), extracted from the ship navigation system. Forecasted wind direction and speed presented in the lower panel are compared to $10$~min average values (corrected for ship motion) acquired by the ship anemometer. The six stops when measurements are performed are clearly visible as when the SOG graph flats out at close to zero value. These periods are highlighted with green background color. \

As the ship cruised between stops, the heading direction was about $190 \degree$. During the measurements, the heading direction was $240 \degree$ at stop $2.1$, roughly $140 \degree$ at stop $2.2$-$2.5$ and about $315 \degree$ at stop $2.6$. The wind direction was almost constantly $45 \degree$ over the six-hour period. It is clear that the ship was oriented approximately perpendicular to the wind direction when it drifted freely during the stops, except from stop $2.1$ where the high ice concentration most likely prevented the ship from rotating. For stop $2.2$-$2.5$, the wind came in on port side of the vessel, while it came in from starboard side at stop $2.6$. \

There is generally a good agreement between forecast and measured wind direction ($WD$) and speed ($WS$), although the model overestimated $WS$ by $50$-$75$\% in the period from roughly 15:30 to 18:30. $THQ$ from model and $H_{S}$ from model and bow measurements are also presented in the lower panel of Fig.~\ref{AWS}. The overestimated wind speed could possibly explain why forecast $H_{S}$ is $95$\% higher than measured value at 20:09 when wave model data were available, as WAM-4 is forced with winds at $10 m$ height from the atmospheric model UM4. Also, occasional ice floes were observed at this point and WAM-4 does not take attenuation caused by the presence of ice into considerations. However, the measured wave data were saturated at this point and cannot be completely trusted. THQ was $282.1 \degree$ at 20:09. Due to the low variation in THQ (in time and space) over the period when attenuation measurements were sampled, as seen in Fig.~\ref{fig:model_outside}, this value ($282.1 \degree$) is used to find the WTD through the MIZ, presented in Table~\ref{ship_stops} and Fig.~\ref{fig:trajectory}. The wave heading angle $\beta$ was $40 \degree$ at stop $2.1$, $140 \degree$ at the four intermediate stops $2.2$-$2.5$ and $35 \degree$ at stop $2.6$. Wave heading angle will be further assessed in Section~\ref{discussion}. \

\begin{figure}
  \begin{center}
    \includegraphics[width=.55\textwidth]{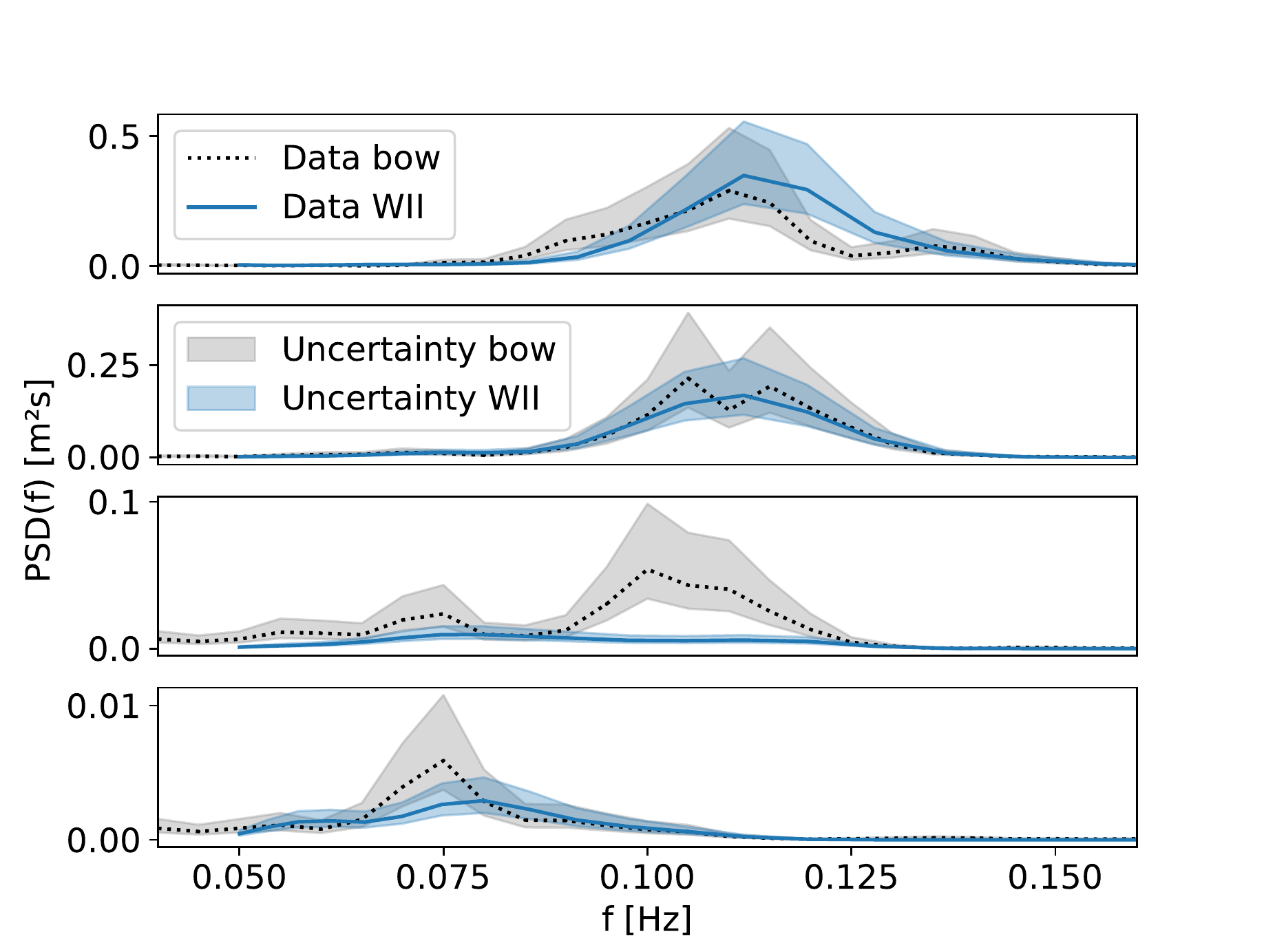} 
    \caption{\label{validation_IMU} Validation of bow measurements in the MIZ with WII instruments placed on ice floes. PSD are presented from bow measurements (black) and from WII instruments (blue) with their respective $95$\% confidence intervals as shaded regions, going further into the ice zone from upper (stop $1.1$) to lower panel (stop $1.4$).}
  \end{center}
\end{figure}

Power spectra from bow measurements (black) and WII instruments (blue) with their respective $95$\% confidence intervals are compared in Fig.~\ref{validation_IMU}, going successively from stop number $1.1$ (upper panel) to $1.4$ (lower panel). The power spectra from the two methods are consistent for most frequencies in the two first samples closer to the open ocean. The third sample is consistent for some frequencies up to $0.09$~Hz. For higher frequencies, the WII spectrum flattens out while the bow spectrum peaks at $0.1$~Hz. Investigation of bow spectra from samples taken respectively twenty and forty minutes after the PSD presented here while the ship was still stationary around the WII instruments, reveals similar spectral peaks at the same frequency. This consistency substantiates the validity of the bow measurements. A discussion on this discrepancy follows in Section~\ref{discussion}. The fourth sample furthest into the ice displays a fair agreement between the two methods. \

Integrated parameters from the spectra in Fig.~\ref{validation_IMU} are summarized in Table~\ref{table:IMU}. There is a good match for both significant wave height and mean period where MPEs do not exceed $\pm10$\% for all stops except $1.3$ where the error in $H_{S}$ is $-87.8$\%. We use MPE as defined in (\ref{MPE}), although it is not strictly a mean error in this context where only single point parameters are compared. \   

\begin{table}[h]
\centering 
\begin{tabular}{c c c c c c c}  
\toprule
Stop & \multicolumn{3}{c}{Significant wave height} & \multicolumn{3}{c}{Mean wave period } \\[0.5ex]
\cmidrule(lr){2-4} 
\cmidrule(lr){5-7}
{} & $H_{S\_bow}$ [m] & $H_{S\_WII}$ [m] & MPE [\%] & $T_{m02\_bow}$ [s] & $T_{m02\_WII}$ [s] & MPE [\%] \\ [0.5ex]
\midrule
1.1 & 0.36 & 0.39 & 8.4 & 8.85 & 8.65 & -2.4 \\[0.5ex] 
1.2 & 0.29 & 0.28 & -4.3 & 9.03 & 9.12 & 1.0 \\[0.5ex] 
1.3 & 0.16 & 0.09 & -87.8 & 10.14 & 10.89 & 6.9 \\[0.5ex]
1.4 & 0.04 & 0.04 & -5.1 & 11.36 & 12.57 & 9.6 \\[0.5ex]
\bottomrule
\end{tabular}
\caption{Significant wave height and zero up-crossing period from bow measurements inside the MIZ are compared with WII measurements. Errors are included. } 
\label{table:IMU} 
\end{table}

\subsection{Wave attenuation} \label{damping_section}

\begin{figure}
  \begin{center}
    \includegraphics[width=.55\textwidth]{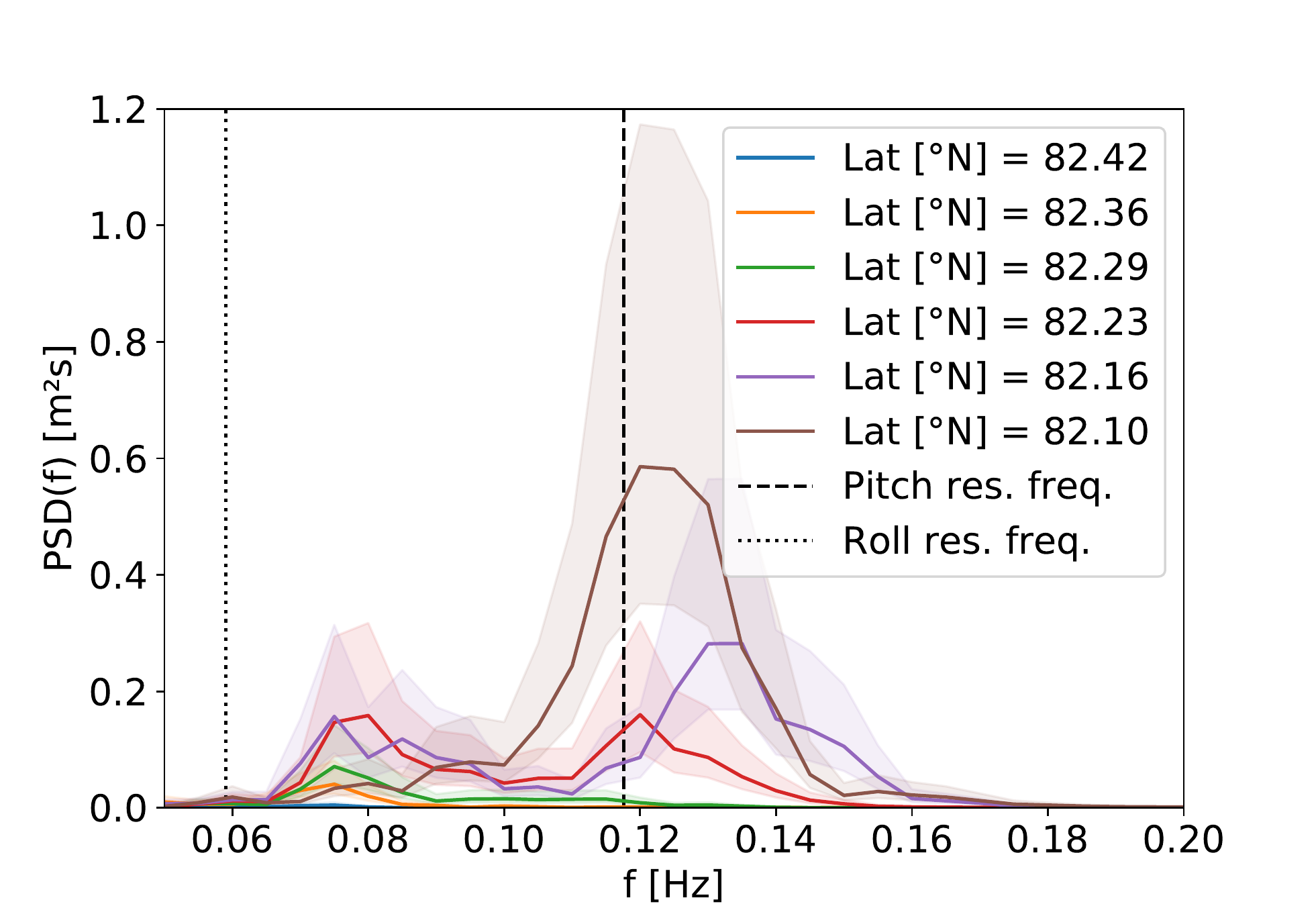} 
    \caption{\label{wave_spectra} Power spectra at different latitudes with $95$\% confidence intervals, and ship resonance frequencies in vertical modes from \cite{ytterland2016motion}.}
  \end{center}
\end{figure}

One dimensional power spectra of the surface elevation with $95$\% confidence intervals from stops $2.1$-$2.6$ are presented in Fig.~\ref{wave_spectra}. Most of the energy is found at either low (swell) or high (wind wave) frequencies. Averaged peak frequency is found to be $0.076$~Hz for swell and $0.128$~Hz for wind wave. Spectral amplitudes $a(f_{0})$ for a set of six finite frequency bins are found from (\ref{a}) and investigated at stops $2.1$-$2.6$. The highest and lowest $f_{0}$ are set to $0.076$~Hz and $0.128$~Hz respectively. Four intermediate forcing frequencies are evenly distributed between the swell and wind wave frequency.  \

\begin{figure}[H]
\begin{subfigure}[H]{0.5\linewidth}
\includegraphics[width=8cm]{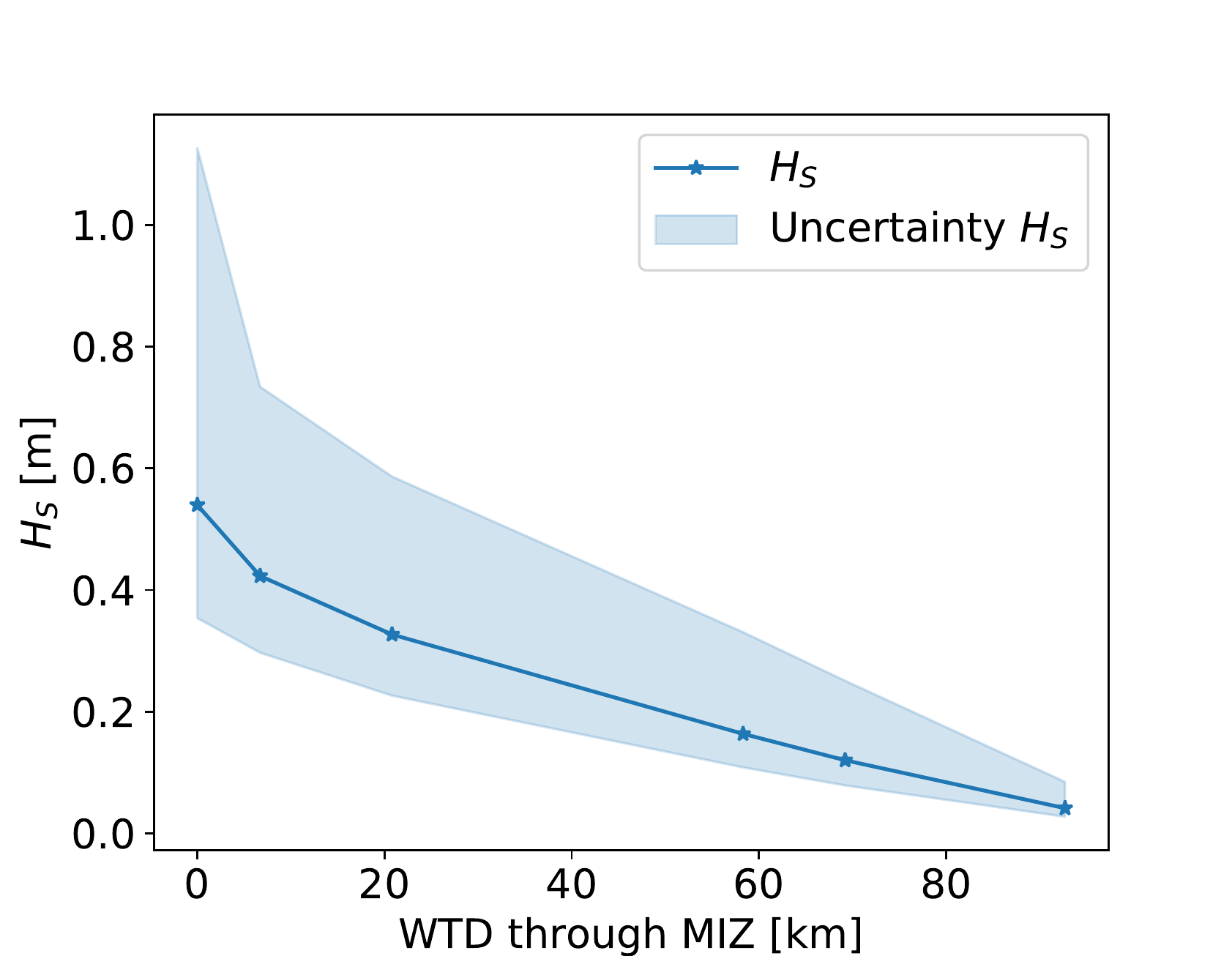} 
\caption{$H_{S}$ versus wave traveling distance into the MIZ.}
\label{fig:Hs}
\end{subfigure}
\hfill
\begin{subfigure}[H]{0.5\linewidth}
\includegraphics[width=8cm]{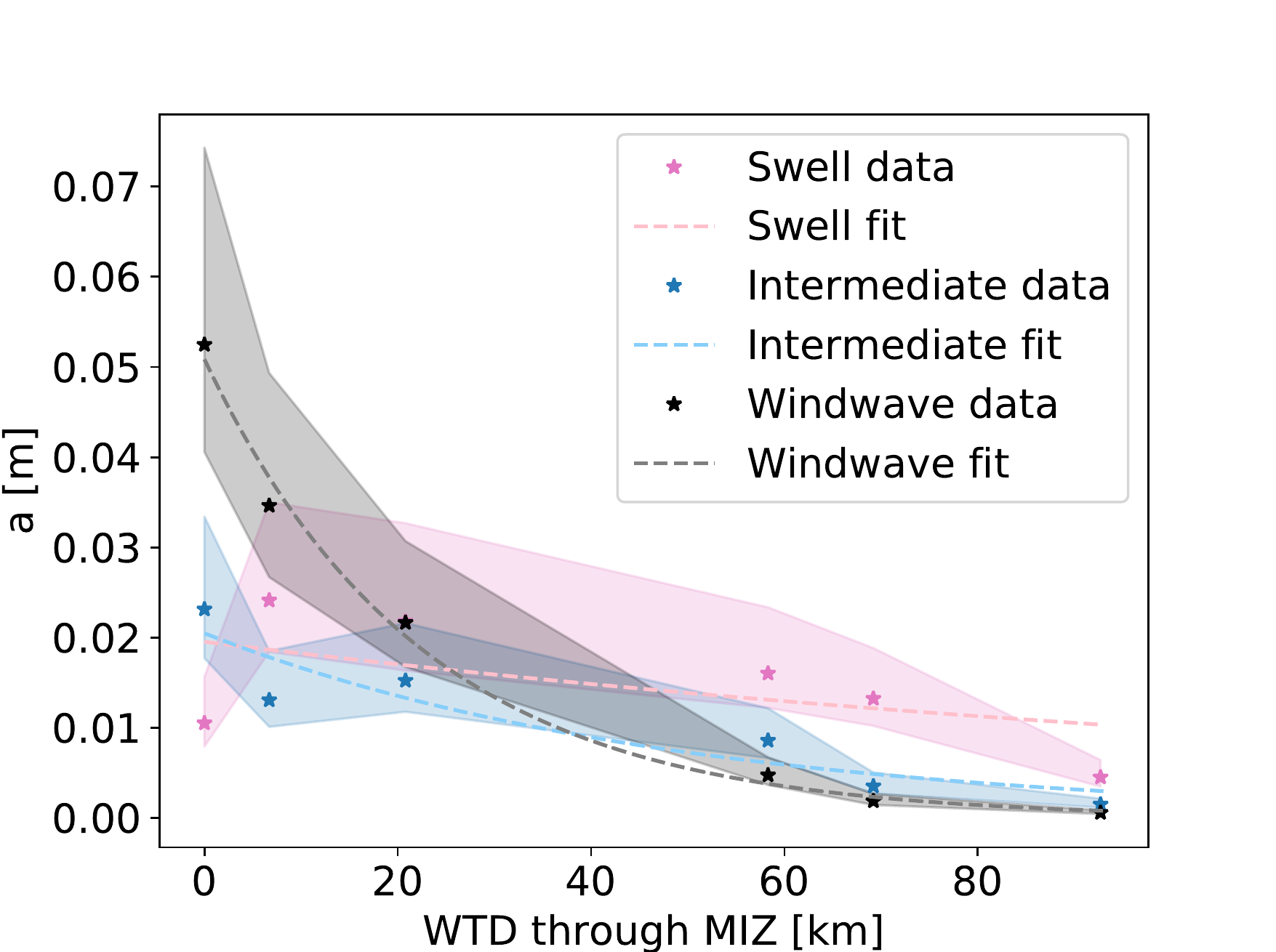} 
\caption{$a(f_{0})$ versus wave traveling distance into the MIZ.}
\label{fig:a}
\end{subfigure}
\hfill
\caption{Wave damping with $95$\% confidence intervals. a) Significant wave height $H_{S}$. b) Spectral amplitudes $a(f_{0})$ for swell, one of the intermediate wave frequencies and wind wave. Exponential decay functions are fitted to the data. }
\label{fig:wave_damping}
\end{figure}

It is evident that wave energy content increases as the ship approaches the open ocean. Figure~\ref{fig:Hs} shows an exponential decrease in $H_{S}$ (with $95$\% confidence intervals) as function of WTD through the MIZ for stops $2.1$-$2.6$. Ideally, the measurements should have been performed simultaneously to ensure stationary wind forcing throughout the attenuation run. Fortunately, wind conditions did not change dramatically over the period from 13:11 to 19:08 as can be seen in Fig.~\ref{AWS}, and we therefore assume the wave generation to be approximately stationary. \

Spectral amplitude for swell, the third intermediate wave, and wind wave are plotted versus wave travel distance through the MIZ and presented in Fig.~\ref{fig:a} along with their $95$\% confidence intervals. A clear exponential attenuation for wind wave can be seen. The exponential shape is less obvious for the intermediate frequency and swell, but the trend is visible. Also, swell amplitude seems to decrease from stop $2.5$ to $2.6$, which is not physical, and this problem is addressed in Section~\ref{discussion}. Exponential fits lie within the confidence intervals of $a(f_{0})$ for almost all frequencies.

\begin{figure}
  \begin{center}
    \includegraphics[width=.55\textwidth]{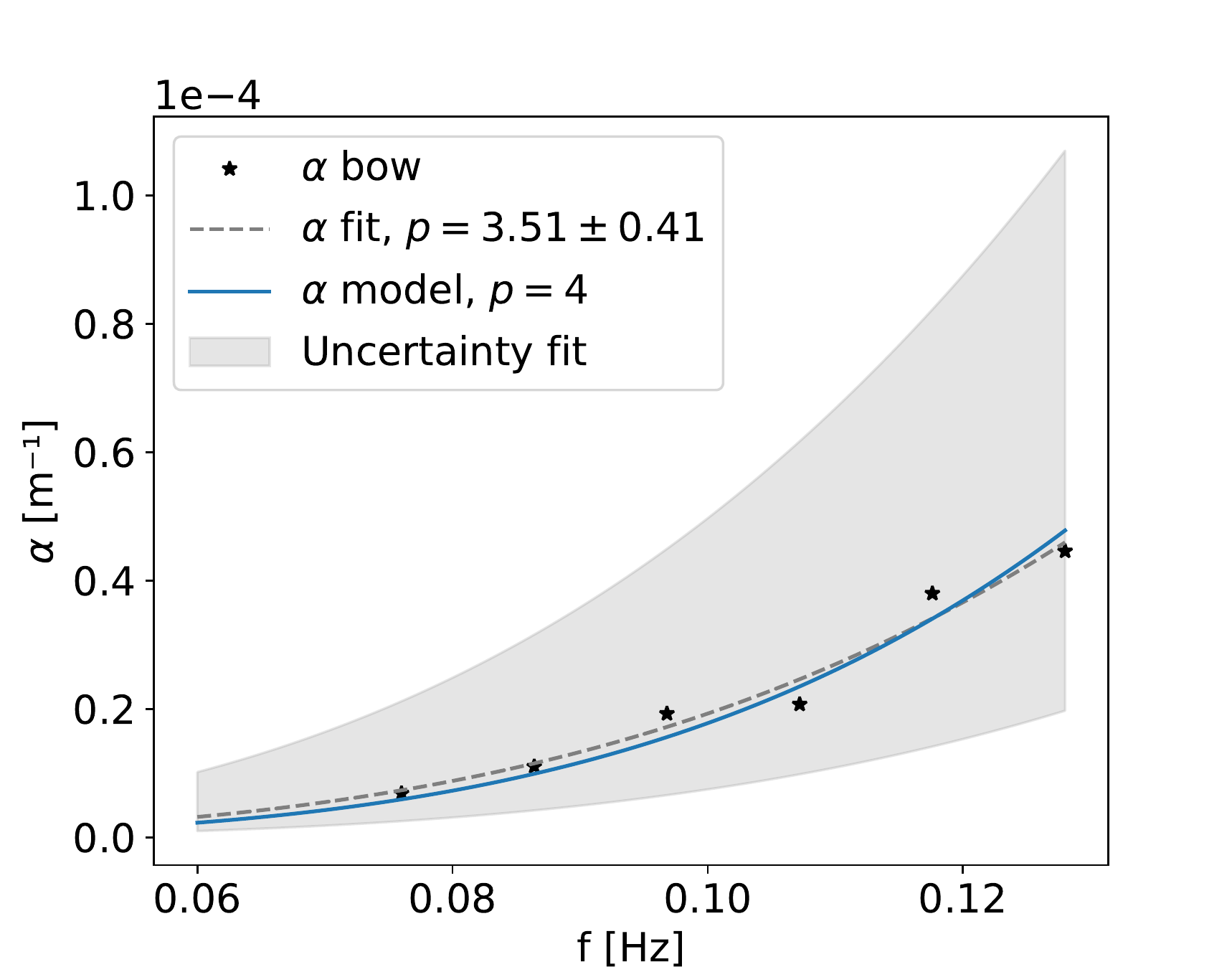} 
    \caption{\label{amplitude_damping} Spatial damping coefficient $\alpha$ as function of frequency from measurements (black) and two-layer model (blue). Power coefficients for the fitted curve and the model are displayed in the legend. $1\sigma$ errors associated with the power coefficient estimate for the fitted curve (gray) are shown as the gray shaded region. }
  \end{center}
\end{figure}

The spatial damping coefficients $\alpha$ are plotted versus their respective $f_{0}$ with black asterisks in Fig.~\ref{amplitude_damping}. We obtain values of $\alpha$ from $0.67\times 10^{-5} \mathrm{m}^{-1}$ at $0.076$~Hz to $4.44\times 10^{-5} \mathrm{m}^{-1}$ at $0.128$~Hz. These values are within the same order of magnitude as other attenuation observations within the same frequencies reported in the literature where the measurements are done over approximately the same wave traveling distance into the MIZ, e.g. \cite{squire1980direct} ($\alpha = 2.7-8.6\times 10^{-5} \mathrm{m}^{-1}$) and the February 26, 1983 event reported in \cite{wadhams1988attenuation} ($\alpha = 1.6-4.8\times 10^{-5} \mathrm{m}^{-1}$). \

Observed wave attenuation increase with frequency in the shape of a power function. The fitted power curve is plotted with a dashed gray line in Fig.~\ref{amplitude_damping}, and the $1\sigma$ uncertainty associated with the power law exponent is marked with a gray shaded area. The fitted exponent is $3.51 \pm 0.41$, where the uncertainty is the standard error from the curve fit. Values of $\alpha$ obtained from (\ref{alpha_two}) are plotted with a blue line in Fig.~\ref{amplitude_damping}. The two-layer model of \cite{sutherland2019two} describes the frequency dependence of $\alpha$ quite well for the frequencies considered here. It can be noted that the model slightly overestimates the frequency dependence for frequencies above $0.12$~Hz. \

\section{Discussion} \label{discussion}

A culture of sharing the design of instruments within the scientific community would enable a more cost effective and increased quantity of data collection in the Arctic. We would like to stimulate researchers to use off-the-shelf sensors and open-source code and therefore make all our material available (See Appendix A). The components used to obtain wave data in this study are relatively cheap compared to industry made black-box instruments. The system can record autonomously, hence, expensive expedition time does not have to be allocated as measurements can be done in parallel. Also, cruises in the Arctic with other purposes than measuring waves in ice, for example studies within biology or meteorology, could benefit from installing the system presented here, which provides wave measurements without the need for human intervention. Other devices such as floaters and buoys offer the advantage of multiple simultaneous measurements at different locations, which is not possible with bow mounted sensors. However, bow mounted sensors do not require costly and time-consuming deployment and retrieval operations, and there is low risk associated with losing instruments at sea. \

The primary focus of this study is to present a new method for wave measurements in the MIZ. Spectral wave models have been included only as a supplement for comparison with the observations. Nevertheless, we provide a short discussion on the difference between the models and how well they correspond to the bow system. As mentioned in Section~\ref{wave_model}, ERA5 is a reanalysis of previous observations and simulations, and is therefore considered more reliable than a forecast model, such as WAM-4. However, WAM-4 matches the observations better than ERA5 in terms of significant wave height, especially in the vicinity of the MIZ. No wave models are the true realization of the ocean, but there are arguments that support the accuracy of WAM-4 over ERA5 in our situation. ERA5 provides wave information for sea ice concentration up to 50\%, while the hard ice boundary in WAM-4 is defined at 30\%. In other words, wave information is available further north from ERA5  than from WAM-4 as seen from Fig.~\ref{validation_WAM}, although none of the models consider the effect of wave damping due to the presence of sea ice. Consequently, the modeled significant wave height furthest into the MIZ deviates the most from the real values. Additionally, ERA5 is a global model with a horizontal resolution of 40~km, while WAM-4 is 10 times finer resolved, as seen in Fig~\ref{fig:satellite}. Considering the fact that the measurements were done relatively close to the Svalbard and Franz Josef Land archipelagos, the horizontal resolution is likely to be more important for the accuracy of the models than for example in the middle of the Pacific. \  

Data recorded while a ship is in motion will contain a Doppler shifted frequency according to the wave heading angle $\beta$ \citep{collins2017doppler}, which could be problematic in the calculation of periods. Peak and mean periods were reported strongly biased by the Doppler shift in \cite{christensen2013surface}, while the estimates of significant wave heights were reasonable since they only depend on the sea surface variance. In the present study, we consider the measurements invalid either when the ship is cruising (and a Doppler shift could be introduced depending on $\beta$) or when the UG is saturated (exceeded measure range), as we have not differentiated between these two cases. We see little difference in MAPE inside versus outside the time periods we consider valid, when comparing periods from bow measurements with model data. MPEs are actually higher in the periods where data is considered valid. There is also a slightly better match outside the valid periods in the comparison of significant wave height. This is somewhat counter intuitive, but there is of course uncertainty associated with the spectral models, and $\beta$ is not included in the analysis. It is possible to correct the Doppler shift, but this requires accurate observations of the wave direction, or more precisely $\beta$ \citep{collins2017doppler}. This would allow for possibly improved mean period estimates during cruising. \ 

Significant wave energy for $f>0.1$~Hz was detected by the bow mounted sensors and not by the WII instruments at stop $1.3$ as seen in Figure~\ref{validation_IMU}. \cite{yiew2016hydrodynamic} have performed experiments with thin floating disks and applied two theoretical models to determine the hydrodynamic responses of ice floes with diameter $d$ in regular waves with wavelength $\lambda$. They find RAO in heave to be $0.5$ at approximately $\lambda/d=1.5$ and rapidly decreasing for smaller $\lambda/d$. For the peak frequency measured by the bow sensor, i.e. $f=0.1$~Hz, $\lambda$ is found to be $156$~m with Eq. \ref{dispersion}. We have no exact measure of the size of the ice floe at stop $1.3$, but deeming from Fig.~\ref{fig:satellite} and Fig. 9 in \cite{rabault2019open} its diameter seems to be at least in the order of hundred meters. The flexural rigidity of the ice floe will of course differ from the disks applied in \cite{yiew2016hydrodynamic}, but the results in this paper suggest a low heave response of such a large floe at this short wavelength. As the WII instruments essentially calculate the power spectra from the heave motion of the floe, a plausible explanation for the discrepancy between the two instruments could be that waves in this frequency range were effectively damped by the floe. The bow sensor on the other hand, measured at points where the ice was broken up by the ship and was therefore able to detect higher frequencies and give a more accurate observation of the sea state. \         

Another relevant problem is the response of the ship itself. Figure~\ref{wave_spectra} shows the resonance frequencies of the ship in the vertical modes with $\beta = 150 \degree/30 \degree$ (due to symmetry) \citep{ytterland2016motion}, which almost corresponds to $\beta$ at stops $2.1$-$2.5$ ($10\degree$ deviation) and stop $2.6$ ($5\degree$ deviation). The RAO values are $1.8$ and $1.9$ for these angles, meaning a possible maximum amplification of $80$\% and $90$\% for roll and pitch motion respectively. Heave mode has only got a peak in RAO for $\beta = 90 \degree$, which did not occur during stops $2.1$-$2.6$ and is therefore not shown. None of the spectral peaks in Fig.~\ref{wave_spectra} coincides with the natural frequencies in roll. However, the high frequency peaks are in the same range as the ship's natural frequency in pitch. Pitch motion was likely amplified up to $90$\% when exposed to $0.12$~Hz waves. However, this effect should be compensated for by the pitch motion correction performed on the data. \

Stop $2.6$ has been included in our wave attenuation analysis even though it was located outside of the defined ice edge in Figure~\ref{fig:trajectory}. This ice edge was estimated from coarsely resolved satellite images and is therefore not very accurate. Reflected wave energy from the ice edge could have been a problem at stop $2.6$. However, the ice edge was not a strictly defined line, it was rather observed as a gradually decreasing ice concentration towards the open ocean and ice floes were still present around stop $2.6$. Swell energy was observed to decrease from stop $2.5$ to $2.6$ where the opposite is expected. Our assumption of stationary wave field may not have been valid in this case, although wind measurements and WAM-4 model data suggest otherwise. There is also uncertainty associated with the total mean wave direction $THQ$. There could have been discrepancies between $THQ$ and mean swell direction $THQ_{swell}$, although $90\degree<THQ_{swell}<270\degree$ (swell traveling south) is unlikely. Another possible explanation is that ship yaw (i.e. rotation in the horizontal plane) and/or a changing $\beta$ might have influenced the low frequencies in the measurements. At stop $2.6$ the ship heading increased continuously from approximately $200\degree$ until it stabilized at $315\degree$ towards the end of the $20$~min sampling period, whereas both ship heading and $\beta$ were almost constant for the other stops, as seen in Figure~\ref{AWS}. \ 

We emphasize that we are not trying to show a universal power law in Fig.~\ref{amplitude_damping}, which would not make sense for frequencies spanning over less than a decade. Our intention is to compare the observed wave attenuation with the two layer waves-in-ice model of \cite{sutherland2019two}, and not to verify the model. The power function curve fit gives a frequency dependence of $\alpha \propto f^{3.51 \pm 0.41}$, which is similar to (\ref{alpha_two}) where $\alpha \propto f^{4}$. Wind energy input is not considered by the model. With a wind speed of approximately $6$~m/s on average over the measurement period as seen in Fig.~\ref{AWS}, it is reasonable to assume that wave generation has taken place, which will reduce the slope of the observed attenuation at higher frequencies. This could explain why the model predicts the observation quite well for the lower frequency bins, while it slightly overestimates attenuation for the higher frequency bin, based on the six data points.

\section{Conclusions} \label{conclusions}

We have presented shipborne wave measurements from the MIZ with a system combining an altimeter (UG) and a motion correction device (IMU). The UG was able to reflect signals off an ice-covered surface. The current system was designed for the present cruise in the MIZ of the Barents Sea where waves are normally quite small. Hence, the choice of the present UG and its auto calibration with a $1$~m range. If some cruise would plan to go to rougher seas, one could either use the same UG over its full range ($8$~m), or another UG with larger range. The methodology is both cost effective as components are off-the-shelf and it can be installed on a cruise with other objectives. The system has proven to be robust as instruments have measured continuously over the two-week period without suffering any damage. Our setup provides single point time series of ocean surface elevation outside of and inside the MIZ, which enables us to produce 1D power spectra and integrated parameters. \

Measured data have been compared to the WAM-4 and the ERA5 spectral wave models over a period of eight days. We have found good agreement in mean and zero up-crossing periods and significant wave height. Mean absolute percentage errors (MAPE) were $18.9$\%, $15.0$\% and $17.2$\% for $H_{S}$, $Tm_{01}$ and $Tm_{02}$ respectively when compared with WAM-4 and $29.2$\% and $13.4$\% for $H_{S}$ and $Tm_{01}$ respectively when compared with ERA5 during valid measurements ($23$\% of the time). Errors were about the same during periods where measurements were not considered valid, indicating that the system is insensitive to exceeded instrument range and/or Doppler shifted wave frequencies. Forecast significant wave height was $95$\% higher than the value measured inside the MIZ. This deviation might have been caused by an overestimated model wind speed compared to wind speed measured with the ship anemometer, or by the fact that occasional ice floes were present, which were not considered by the forecast model. Power spectra and integrated parameters from bow measurements have been compared with values obtained from waves-in-ice instruments placed on ice floes. A good match was found when the sensors were placed on smaller floes further out in the MIZ. The spectra agree fairly well further into the MIZ. A major discrepancy was likely caused by a large ice floe that filtered out higher frequencies where the waves-in-ice instrument was placed. Errors in $H_{S}$, $Tm_{02}$ were smaller than $10$\% in all cases except for one of the measurements. \

We observed an exponential wave attenuation when going through the MIZ. The spatial damping coefficients obtained from measurements were within the same order of magnitude as observations reported in previous studies with a similar wave traveling distance through the ice \citep{squire1980direct,wadhams1988attenuation}. We have found a strongly frequency dependent attenuation, where $\alpha \propto f^{3.51 \pm 0.41}$, which is similar to a two-layer attenuation model where $\alpha \propto f^{4}$ \citep{sutherland2019two}. \

\section*{Acknowledgement}

The authors are grateful to Øyvind Breivik for inviting us to the cruise. We also thank the crew of RV Kronprins Haakon for their assistance and Yurii Batrak at MET Norway for providing satellite images. The Nansen Legacy project helped funding the cruise and loggers. Funding for the experiment was provided by the Research Council of Norway under the PETROMAKS2 scheme (project DOFI, Grant number $28062$). The data are available from the corresponding author upon request.

\section*{Appendix A: Open source code and designs} \label{appendix:a}

All the designs and files used for setting up the system, including the code used to extract and process data, and general instructions for mounting the instruments, are made available on the Github of the author under a MIT license that allows full re-use and further development (\url{https://github.com/jerabaul29/Ultrasound_IMU_boat_waves_system} [Note: The material will be available upon publication in peer-reviewed literature]). All software and designs are based entirely on open source tools, so that the designs can be easily modified and built upon.

\begin{footnotesize}
\bibliographystyle{agsm} 
\bibliography{proposal_v3.bib}
\end{footnotesize}

\end{document}